\documentclass[prd,preprint,superscriptaddress,preprintnumbers,eqsecnum,showpacs,nofootinbib,nobibnotes]{revtex4}
\usepackage{amsfonts,amsmath,amssymb,bm,natbib}
\usepackage{graphicx} 

\newcommand{\be}{\begin{equation}}
\newcommand{\bea}{\begin{eqnarray}}
\newcommand{\ee}{\end{equation}}
\newcommand{\eea}{\end{eqnarray}}

\def\Y{Y}

\def\spr{\!\cdot\!}

\def\s#1{{\scriptscriptstyle #1}}

\def\noeq#1{(\ref{#1})}
\def\1eq#1{Eq.~(\ref{#1})}

\def\2eqs#1#2{Eqs.~(\ref{#1}) and~(\ref{#2})}
\def\3eqs#1#2#3{Eqs.~(\ref{#1}),~(\ref{#2}) and~(\ref{#3})}

\def\fig#1{Fig.~\ref{#1}}

\def\ie{{\it i.e.}, }
\def\eg{{\it e.g.}, }

\def\n#1{({\it #1}\,)}

\def\cd{\!\cdot\!}

\begin{document}

\title{Effects of divergent ghost loops on the Green's functions of QCD}

\author{A.~C. Aguilar}
\affiliation{University of Campinas - UNICAMP, 
Institute of Physics ``Gleb Wataghin'' \\
13083-859 Campinas, SP, Brazil}

\author{D. Binosi}
\author{D. Iba\~nez}
\affiliation{European Centre for Theoretical Studies in Nuclear
Physics and Related Areas (ECT*) and Fondazione Bruno Kessler, \\Villa Tambosi, Strada delle
Tabarelle 286, 
I-38123 Villazzano (TN)  Italy}

\author{J. Papavassiliou}
\affiliation{\mbox{Department of Theoretical Physics and IFIC, 
University of Valencia and CSIC},
E-46100, Valencia, Spain} 

\begin{abstract}
In the present work we discuss certain characteristic features encoded in some of the fundamental  QCD Green's functions, whose origin can be traced back to the nonperturbative masslessness of the ghost field, in the  Landau gauge. Specifically,  the ghost  loops that  contribute to
these Green's  functions display  infrared divergences, akin  to those
encountered in the perturbative treatment, in contradistinction to the
gluonic  loops,  whose  perturbative  divergences  are  tamed  by  the
dynamical  generation  of an  effective  gluon  mass.   In $d=4$,  the
aforementioned divergences are  logarithmic, thus causing a relatively
mild impact, whereas in $d=3$ they are linear, giving rise to enhanced
effects.  In  the case of the  gluon propagator, these  effects do not
interfere with  its finiteness, but make its  first derivative diverge
at  the origin,  and introduce  a maximum  in the  region  of infrared
momenta.  The three-gluon  vertex is  also affected,  and  the induced
divergent  behavior is  clearly exposed  in certain  special kinematic
configurations, usually considered in lattice simulations; the sign of
the  corresponding divergence  is unambiguously  determined.  The main
underlying  concepts are  developed in  the  context of  a simple  toy
model,  which demonstrates  clearly the  interconnected nature  of the
various   effects.    The  picture   that   emerges  is   subsequently
corroborated by a detailed nonperturbative analysis, combining lattice
results with  the dynamical integral equations  governing the relevant
ingredients,  such   as  the   nonperturbative  ghost  loop   and  the
momentum-dependent gluon mass.
\end{abstract}


\maketitle

\section{Introduction}

In recent years our understanding of the infrared (IR) sector of QCD has advanced considerably, 
due to a detailed and systematic scrutiny of the fundamental Green's functions of the theory. 
In particular, high quality lattice simulations of propagators~\cite{Cucchieri:2007md,Cucchieri:2010xr,Bogolubsky:2009dc,Oliveira:2009eh,Ayala:2012pb}  
and vertices~\cite{Cucchieri:2006tf,Cucchieri:2008qm} have furnished new insights on the subtle underlying mechanisms, and have spurred 
an intense  parallel activity within the various nonperturbative approaches 
in the continuum~\cite{Boucaud:2008ky,Aguilar:2008xm,Fischer:2008uz,RodriguezQuintero:2010wy,Pennington:2011xs,Campagnari:2010wc,Alkofer:2000wg,Maris:2003vk,Aguilar:2004sw,Fischer:2006ub,Kondo:2006ih,Braun:2007bx,Binosi:2007pi,Binosi:2008qk,Kondo:2011ab,Szczepaniak:2001rg,Szczepaniak:2003ve,Epple:2007ut,Szczepaniak:2010fe,Watson:2010cn,Watson:2011kv}.

At  this point,  the plethora  of  available information  needs to  be
interpreted carefully, and  be used in the construction  of a reliable
picture  of  the  fundamental  dynamics,  with  increasingly  stronger
predictive power.  To that  end, in the  present work we  elaborate on
what appears to  be a profound connection between  the masslessness of
the  ghost, the  precise  form of  the  gluon propagator  in the  deep
IR, and the divergences  observed in certain kinematic limits of
the  three-gluon vertex. This  particular connection  is valid  in the
Landau  gauge,  both in  $d=3,4$;  however,  in  $d=3$ the  associated
effects are  considerably more enhanced, for reasons  that will become
clear in what follows.

As is well-known by now, 
the infrared finiteness of the gluon propagator and the ghost dressing function, observed in 
a variety of (Landau gauge) lattice simulations, 
may be explained in a rather natural way by invoking the concept of a 
dynamically generated mass~\cite{Cornwall:1981zr,Bernard:1982my,Donoghue:1983fy,Philipsen:2001ip}. 
In particular, the (Euclidean) gluon propagator $\Delta(q^2)$ 
assumes the form $\Delta^{-1}(q^2) =q^2 J(q^2) + m^2(q^2)$,  
where the first term corresponds to the ``kinetic term'', or ``wave function'' contribution, 
while the second denotes the momentum-dependent mass function~\cite{Aguilar:2011ux,Binosi:2012sj}. 
Within the framework of the Schwinger-Dyson equations (SDEs) 
both $J(q^2)$ and $m^2(q^2)$ satisfy two independent but coupled integral equations, which, 
at least in principle, determine their dynamical evolution.  

In $d=4$, the main observation underlying the present work may be described as follows. 
The fact that the ghost propagator, $D(q^2)$, remains massless, has as consequence that the contribution to $J(q^2)$ 
stemming from the ghost-loop diagram [$(a_3)$ in~\fig{QB-SDE}] contains a  pure logarithm, $\ln q^2$, which is ``unprotected'', 
in the sense that there is no mass term in its argument that could tame its divergence in the IR. 
This is to be contrasted with the corresponding logarithms originating from the gluonic loops [$(a_1)$ in~\fig{QB-SDE}], 
of the type $\ln (q^2+m^2)$, which, due to the presence of the dynamical gluon mass $m^2(q^2)$, are finite for arbitrary Euclidean momenta. 
Of course, the massless logarithm does not interfere with the overall finiteness of $\Delta(q^2)$, 
simply because it is multiplied by $q^2$; 
its presence, however, makes the first derivative of $\Delta(q^2)$ diverge at the origin. In addition, 
it induces a subtle effect on the  precise shape of the gluon propagator in the deep IR. Specifically, $\Delta(q^2)$ 
is not a monotonic function of $q^2$, displaying a (numerically small) maximum, precisely due to the $q^2\ln q^2$ term. 
The size and location of this effect is largely controlled by the relative weight with which the 
two types of logarithm contribute to $J(q^2)$; in particular, 
the weight of the massless logarithm is about one order of magnitude less than that of the massive, 
a fact which pushes the appearance of the effect in the deep IR, reducing at the same time its size.      

It turns out that the quantity which accounts for the divergent behavior of the  three-gluon vertex (for recent studies of this vertex see also~\cite{Huber:2012kd,Pelaez:2013cpa}) in some  special kinematic limits 
studied on the lattice, is precisely the $J(q^2)$ considered above. 
In particular, in the ``orthogonal'' configuration with one momentum vanishing, 
the usual quantity employed 
in the lattice studies, to be denoted by $R$,
satisfies $R(q^2)\sim[q^2 J (q^2)]'$, where the ``prime'' denotes derivative with respect to $q^2$. 
Thus, the dominant contribution as $q^2\to 0$ is  \mbox{$R(q^2)\sim J (q^2) \sim  \ln q^2$}; evidently, 
for sufficiently small  $q^2$, $R(q^2)$ becomes negative, and diverges as a logarithm.

In $d=3$, the situation is  qualitatively similar to the one described
above,  but the  divergences induced  due to  the masslessness  of the
ghost are stronger. Specifically, as may be already established at the
level of  a simple one-loop  calculation~\cite{Aguilar:2010zx}, the part of  $J(q^2)$ coming
from the ghost loop behaves like $1/q$. As a result, the corresponding
effects  are   significantly  enhanced:  the  maximum   of  the  gluon
propagator is  clearly visible  on the lattice~\cite{Cucchieri:2010xr},  and so is  the abrupt negative divergence seen in the corresponding $R(q^2)$~\cite{Cucchieri:2006tf}.

Note that our  theoretical prediction for the signs  of the divergence
both at  $d=3,4$ is unequivocal: they are fixed by the sign of the logarithm obtained from 
graph $(a_3)$ in~\fig{QB-SDE}  (for earlier related works, see,
\eg~\cite{Schleifenbaum:2006bq,Huber:2007kc}).
In addition,  it is interesting to note that the  observed divergences
occur  within  a   theory  with  a  finite  gluon   propagator  and  a
non-enhanced  ghost dressing  function.  In fact,  the  origin of  the
divergences encountered in the three-gluon vertex is not associated in
any way with the  (intrinsically divergent) ``scaling'' solutions~\cite{Alkofer:2000wg,Fischer:2006ub}, but
rather with the loop effects of massless (but non-enhanced) ghosts.

The article is organized as follows. In Section~\ref{toy} we present a simple description of the gluon propagator,  
which captures quite faithfully all qualitative features mentioned above. This section serves as a reference 
for fixing the main ideas, and can guide the reader through the more complex analysis that follows.  In Section~\ref{fullnon} 
we venture into the full nonperturbative analysis of the divergent ghost loop, and the implication 
for the gluon propagator and the three-gluon vertex. Throughout this 
study we make extensive use of the full nonperturbative equation governing the momentum evolution of the 
gluon mass, derived in~\cite{Binosi:2012sj}. 
Finally, in Section~\ref{concl} we discuss our main results and present our conclusions. 
The article ends with two Appendices, one  where the $R$-projector is discussed in 
a technically simplified but qualitatively accurate setting, and one where the subleading nature of the transverse 
part of the ghost-gluon vertex is established.

\section{Massive versus massless loops: A qualitative description}\label{toy}

In this section we discuss the general ideas that underly the present work, 
and introduce a simple, one-loop inspired model, which explains, with little calculational effort, 
the main effects.  

\subsection{General considerations}
In what follows we will work in the Landau gauge, where the full gluon propagator takes the form
\be
i\Delta_{\mu\nu}(q)=-iP_{\mu\nu}(q)\Delta(q^2);\qquad 
P_{\mu\nu}(q)=g_{\mu\nu}-q_\mu q_\nu/q^2.
\ee
In addition, the ghost propagator $D(q^2)$ and its dressing function, $F(q^2)$, are related by 
\be
D(q^2)= \frac{F(q^2)}{q^2}.
\label{ghostdress}
\ee

\begin{figure}[!t]
\includegraphics[scale=1.1]{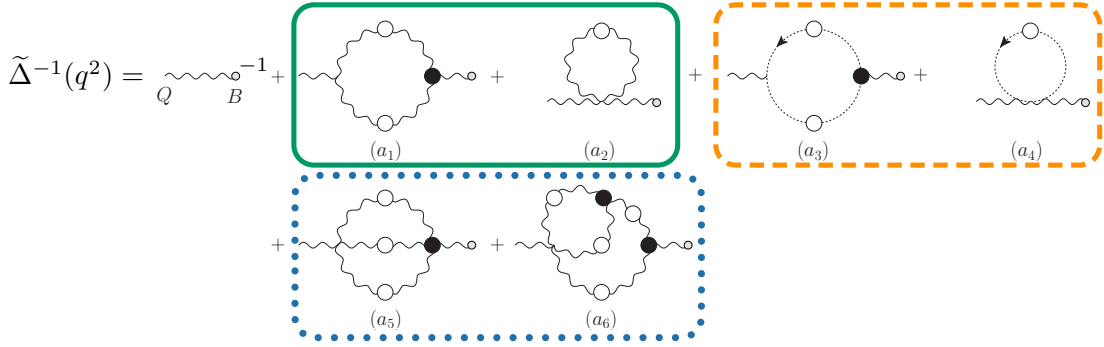} 
\caption{\label{QB-SDE}(color online). The SDE obeyed by the $QB$ gluon propagator. Each of the three different boxes (continuous, dashed, and dotted line) encloses a set of diagrams forming a gauge-invariant subgroup. 
Black (white) blobs represent fully dressed 1-PI (connected) Green's functions; finally, small gray circles appearing on the external legs indicate  background gluons.}
\end{figure}

We will now consider the SDE 
obtained through the combination of the pinch technique (PT)~\cite{Cornwall:1981zr,Cornwall:1989gv,Binosi:2002ft,Binosi:2003rr,Binosi:2009qm}  
with the background field method (BFM)~\cite{Abbott:1980hw},  
known as the PT-BFM scheme~\cite{Aguilar:2006gr,Binosi:2007pi,Binosi:2008qk}.
Specifically, the SDE for the conventional gluon propagator reads 
\be
\Delta^{-1}(q^2){ P}_{\mu\nu}(q) = 
\frac{q^2 {P}_{\mu\nu}(q) + i\,\sum_{i=1}^{6}(a_i)_{\mu\nu}}{1+G(q^2)},
\label{sde}
\ee
where the diagrams $(a_i)$ are shown in Fig.~\ref{QB-SDE}. 
Note that these diagrams give rise to the self-energy of $\widetilde\Delta(q^2)$, namely 
the propagator formed by a quantum gluon ($Q$) and a background one ($B$). 
Thus, \1eq{sde} is the nonperturbative diagrammatic representation of the formal relation 
\be
[1+G(q^2)] \Delta^{-1}(q^2) = \widetilde\Delta^{-1}(q^2), 
\label{conv}
\ee
known in the literature~\cite{Grassi:1999tp,Binosi:2002ez} as a Background-Quantum identity (BQI). 
A SDE similar to that of \1eq{sde}, but with more diagrams, relates $\Delta(q^2)$ with the 
propagator $\widehat\Delta(q^2)$, formed by two background gluons ($B^2$)~\cite{Aguilar:2006gr}; the corresponding BQI reads
\be
[1+G(q^2)]^2 \Delta^{-1}(q^2) = \widehat\Delta^{-1}(q^2). 
\label{conv2}
\ee
The auxiliary function $G(q^2)$ has been studied in detail in~\cite{Aguilar:2009pp}; here it should suffice to 
mention that, for practical purposes, throughout the present work we will use the approximate  
relation 
\be
 1 + G(q^2) \approx F^{-1}(q^2),
\label{GFapp}
\ee
which becomes exact in the deep IR, in $d=3,4$~\cite{Grassi:2004yq,Aguilar:2009nf,Aguilar:2009pp,Aguilar:2010gm}.

As was already mentioned in the Introduction, in the case of an IR finite gluon propagator, 
the scalar function $\Delta(q^2)$ can be decomposed as (Euclidean space)
\be
\Delta^{-1}(q^2)=q^2J(q^2)+m^2(q^2),
\label{prop-deco}
\ee
where $J(q^2)$ is the inverse of the gluon dressing function and $m^2(q^2)$ is 
the dynamically generated (momentum dependent) gluon mass. 
Note that~\1eq{conv} is satisfied 
separately by the kinetic and the mass terms~\cite{Aguilar:2011ux}; thus, using the approximation~\noeq{GFapp}, we have
\be
J(q^2)=F(q^2)\widetilde{J}(q^2);\qquad m^2(q^2)=F(q^2)\widetilde{m}^2(q^2).
\label{indiv-conv}
\ee
A completely analogous relation, 
obtained from \1eq{conv2}, relates $J(q^2)$ with $\widehat{J} (q^2)$ [see \1eq{JhatJ}], 
as well as the corresponding gluon masses.

Given that the function $F(q^2)$ has been simulated accurately on the lattice, 
\1eq{indiv-conv} allows one to obtain $J(q^2)$ from $\widetilde{J}(q^2)$; the latter  
is easier to calculate, due to the special properties of its diagrammatic expansion, 
implemented by the PT-BFM Feynman rules. In particular, 
we remind the reader that all subsets of graphs enclosed within each box~\fig{QB-SDE} give rise 
to a transverse contribution~\cite{Aguilar:2006gr,Binosi:2008qk}; thus, their 
individual treatment (or the total omission of the ``two-loop dressed'' subset) does not compromise the 
transversality of the gluon self-energy.

Even though the dynamical equation governing $\widetilde{J}(q^2)$ [or, equivalently, $J(q^2)$]
is not fully known, mainly due to the poor knowledge of the four-gluon vertex appearing in the 
``two-loop dressed'' diagram of $(a_5)$, the main effect that we want to study 
here originates from the two sets of ``one-loop'' dressed graphs,
namely $(a_1)+(a_2)$ and $(a_3)+(a_4)$.  

It turns out that there is a profound qualitative difference between these two sets of graphs, 
which manifests itself in the behavior of the resulting  $J(q^2)$. 
Specifically, the corresponding contributions to $J(q^2)$ reflect the 
fact that the virtual particles forming these loops (gluons and ghosts, respectively) 
have completely different 
behavior in the IR: while the gluons are effectively massive, the ghosts behave  as massless particles, $D(q^2)\sim1/q^2$.
As a result, in $d=4$, whereas the perturbative logarithm emerging from the first set of graphs 
is tamed by the presence of the gluon mass,
and is therefore finite for all momenta,  
the corresponding logarithm coming from the ghost loop 
remains massless, and, as a consequence, it 
vanishes at a finite value of $q^2$, then reverses its sign, becoming finally divergent at $q^2=0$.   
A similar situation occurs in the $d=3$ case, but the corresponding divergences are linear in $q$ 
instead of logarithmic.

\subsection{The toy model}

The picture described above may be concisely captured by setting 
\be
J_{a_1}(q^2)\sim\left\{
\begin{array}{ll}
\ln\left[(q^2+m^2)/\mu^2\right], & \quad d=4; \\
(1/q)\arctan(q/2m), & \quad d=3,
\end{array}
\right.
\ee 
and 
\be
J_{a_3}(q^2)\sim\left\{
\begin{array}{ll}
\ln\left(q^2/\mu^2\right), & \quad d=4; \\
1/q, & \quad d=3.
\end{array}
\right.
\label{modJ}
\ee 
The corresponding gluon propagator then becomes 
\bea
\Delta^{-1}(q^2) &=& q^2J(q^2)+m^2
\nonumber\\
&=& q^2[1 + c_1J_{a_1}(q^2)+c_2J_{a_3}(q^2)] + m^2,
\label{tm-prop}
\eea
with $c_1$ and $c_2$ two real constants, whose values will be fixed according to arguments given below. 

In the case of $d=4$, the form proposed for $J_{a_3}(q^2)$ corresponds 
simply to the one-loop integral $\int\frac1{k^2(k+q)^2}$, 
reflecting the fact that the internal ghost propagators are massless. 
On the other hand, $J_{a_1}(q^2)$ simulates an integral whose internal propagators are massive\footnote{We hasten to emphasize that we do not advocate the use 
of naive massive gluons inside loops as a self-consistent 
theoretical option.
In fact, such an approach would clash 
with a number of field-theoretic principles that the PT-BFM formalism is designed to preserve, 
such as the transversality of the gluon self-energy.}.
As a result, the subset of logarithmic contributions originating from gluon loops [practically $(a_1)$ in~\fig{QB-SDE}]
undergoes the replacement\footnote{A loop with hard masses gives rise to the text-book integral $\int_0^1 {\rm d}x \ln [q^2 x(1-x) + m^2]$; 
however, the resulting expression does not provide any further insights to the question at hand 
than the simple massive logarithm employed here.}
$\ln\left(q^2/\mu^2\right) \to \ln\left[(q^2+m^2)/\mu^2\right]$. The presence of the 
mass prevents this logarithm from diverging; depending on the ratio $m/\mu$, the logarithm may turn 
negative past a certain value of $q^2$, but remains finite, reaching the final value 
$\ln\left(m^2/\mu^2\right)$.

On the other hand, in the $d=3$ case the corresponding transition from massless to massive loops
is implemented through the substitution (Minkowski space)
\be
\int_k\frac1{k^2(k+q)^2}=\left(\frac{i}{8}\right) \frac1{q} \longrightarrow 
\int_k\frac1{(k^2-m^2)[(k+q)^2-m^2]} =\left(\frac{i}{4\pi}\right)\frac1{q}\arctan\left(\frac{q}{2m}\right).
\ee

Returning to the values of $c_1$ and $c_2$, let us first focus on the $d=4$ case. 
Given that the proposed toy model is clearly one-loop inspired,
it is natural to expect that the values of $c_1$ and $c_2$ would be determined 
by the pre-factors multiplying the corresponding one-loop diagrams. Specifically, one has
\be
c_1 = 2 \left(\frac{\alpha C_A}{4\pi}\right);\qquad c_2=  \frac{1}{6}\left(\frac{\alpha C_A}{4\pi}\right),
\ee
where $C_A$ is the Casimir eigenvalue in the adjoint representation [$C_A=N$ for SU$(N)$], and \mbox{$\alpha_s =g^2/4\pi$}.
Note that $c_1 + c_2 = \frac{13}{6}\left(\frac{\alpha C_A}{4\pi}\right)$, which is precisely the 
total coefficient appearing in the well-known one-loop result~\cite{Pascual:1984zb}, 
$\Delta^{-1}(q^2) =1 + \frac{13}{6} q^2\left(\frac{\alpha C_A}{4\pi}\right) \ln\left(q^2/\mu^2\right)$.
In obtaining these values 
we have used the asymptotic (ultraviolet) one-loop result 
\mbox{$F(q^2) = 1  - \frac{3}{4}\left(\frac{\alpha C_A}{4\pi}\right) \ln\left(q^2/\mu^2\right)$},
and have replaced the perturbative logarithm by a massive one, since, as mentioned already, 
nonperturbatively the function $F(q^2)$ saturates at a finite value.  

The corresponding values for  $c_1$ and $c_2$ in $d=3$ 
may be determined following a completely analogous procedure, using certain 
auxiliary results presented in~\cite{Aguilar:2010zx}; they are given by 
\be
c_1 =  - \left(\frac{25 g^2 C_A}{32\pi}\right);\qquad c_2= - \left(\frac{g^2 C_A}{32}\right) 
\ee
(notice that there is no $\pi$ in $c_2$).  In addition, since in $d=3$ the gauge coupling  $g^2$ has 
dimensions of mass, so do  $c_1$ and $c_2$. 

In the analysis that follows we will depart from these particular values of $c_1$ and $c_2$, in order to 
expose better the underlying effects. The main lessons that we will retain from the 
one-loop discussion given above are : \n{i} in $d=4$, both $c_1$ and $c_2$ are positive, 
\n{ii} in $d=3$, both $c_1$ and $c_2$ are negative, and \n{iii}
in both cases, $c_2$ is significantly smaller than $c_1$. 

\subsection{The main implications}
 
The model presented above leads to important consequences for the gluon propagator  
and for the three-gluon vertex. 

\subsubsection{The maximum of the gluon propagators}

The gluon propagator, $\Delta(q^2)$, of this toy model displays a maximum, both in $d=3,4$, or, equivalently, 
the inverse propagator, $\Delta^{-1}(q^2)$, displays a minimum.
This can be easily established by taking the first derivative of~\1eq{tm-prop};
specifically, in $d=4$ (and with $m^2$ constant)
\bea
[\Delta^{-1}(q^2)]^{\prime} &=& [q^2 J(q^2)]^{\prime}
\nonumber\\
&=& c_2 \ln\left(q^2/\mu^2\right) + \left\{ 1+ c_1\ln\left[(q^2+m^2)/\mu^2\right] +  \frac{c_1 q^2}{q^2+m^2} +c_2   \right\}. 
\label{der4}
\eea
The quantity in curly brackets is finite, but in general of indefinite sign, due to the presence of the 
logarithm. However, it is clear that for $q^2> \mu^2$ it is positive definite, and so is the massless logarithm; 
in fact, for $q^2 \gg m^2$ the two logarithms combine to give the asymptotic result 
$\frac{13}{6} \left(\frac{\alpha C_A}{4\pi}\right) \ln\left(q^2/\mu^2\right)$. On the other hand, in the opposite momentum limit, 
since the first term can become arbitrarily large and negative as $q^2$ approaches zero (remember, $c_{1,2}>0$), 
there exists a value $0< q^2_\s{\Delta}<\mu^2 $ such that $[\Delta^{-1}(q^2_\s{\Delta})]^{\prime} = 0$, no matter
how small $c_2$ may be; of course, as $c_2$ assumes smaller values,  $q^2_\s{\Delta}$ is pushed closer to zero.
It is then elementary to show that the above zero of the derivative corresponds to a minimum of $\Delta^{-1}(q^2)$, since 
the second derivative is positive, 
\be
[\Delta^{-1}(q^2)]^{\prime\prime} = \frac{c_1}{q^2+m^2} + \frac{c_1 m^2}{(q^2+m^2)^2} + \frac{c_2}{q^2} > 0.
\ee
Thus, one reaches the conclusion that the IR divergence of the first term, 
caused by the massslessness of the ghost, 
and the positivity of the ultraviolet logarithms, reflecting the asymptotically free nature of the theory, 
force $\Delta(q^2)$ to have a maximum.
In what follows we will denote the location of this maximum by $q_{\s \Delta}$.

Let us emphasize that the above conditions are {\it sufficient} but not necessary for the existence of such a maximum.
Indeed, one can easily imagine eliminating the divergent logarithm, by setting $c_2=0$, or saturating it (artificially) 
with some mass. Then, even though everything is finite in the IR, depending on the 
relative values of parameters and masses, one {\it may} still get the rhs of \1eq{der4} to vanish. 
But, if the  massless logarithm is there, \1eq{der4} will always have a solution.
 
It is of course obvious that the quantity $[q^2 J(q^2)]^{\prime}$ 
displays a minimum located exactly at the same point where the maximum of $\Delta(q^2)$ is, and that  the reason for this 
coincidence is simply the constancy of the gluon mass.  
However, in anticipation of the full nonperturbative analysis of the next section, 
where the gluon mass will be a function of $q^2$, we will 
already at this level distinguish these two points, 
by introducing a different symbol 
for  the location of the minimum of the kinetic term, namely $q_{\s J}$.
So, whereas within the toy model we have trivially $q_{\s \Delta}=q_{\s J}$, 
in the complete nonperturbative treatment we will have $q_{\s \Delta} \neq q_{\s J}$.

An analogous proof may be constructed for the $d=3$ case, where the corresponding differentiation yields  
\be
[\Delta^{-1}(q^2)]^{\prime} =[q^2 J(q^2)]^{\prime}
=1 +  \frac{c_1}{2q} \arctan(q/2m) + \frac{c_2}{2q} + \frac{c_1 m}{q^2+4m^2}.
\label{der3}
\ee
To find the approximate $q_\s{\Delta}$ for which the rhs vanishes, 
assume that  $q < m$, expand, and keep only first order terms in $q/m$. Then,
one arrives at the simple solution
\be
q_\s{\Delta}/m = -\frac{c_2/m}{2+ c_1/m}.
\ee
For this solution to be consistent, we must have $0<q_\s{\Delta}/m <1$, or (remember that $c_{1,2} <0$) $\vert c_1\vert+\vert c_2\vert<2m$. This, in turn, restricts the allowed values of the (dimensionless) ratio $m/2g^2$; in particular, at the one-loop level  our toy model provides the lower bound
\be
\frac{m}{2g^2}\gtrsim 0.14,
\ee 
in agreement with a plethora of independent studies based on a variety of approaches 
in the continuum~\cite{Alexanian:1995rp,Buchmuller:1996pp,Eberlein:1998yk,Karabali:1998yq} and on the lattice\footnote{A review on the subject of $d=3$ Yang-Mills theories can be found in~\cite{PTbook}.}~\cite{Karsch:1996aw,Cucchieri:2001tw,Heller:1997nqa,Nakamura:2003pu}.

\subsubsection{The negative divergence of the three-gluon vertex}

In order to understand how the (negative) IR divergences that appear in the studies of special kinematic configurations of the conventional three-gluon vertex ($Q^3$)
are related to the properties of the $J(q^2)$, 
it is convenient to consider a model inspired by the PT-BFM three-gluon vertex ($B^3$)~\cite{Cornwall:1989gv,Binger:2006sj,Ahmadiniaz:2012xp}, 
described in detail in Appendix~\ref{appA}. 
The treatment of the conventional three-gluon vertex will be 
addressed in the next section; it basically boils down to a technically more involved realization of the main idea presented here.

The usual quantity employed 
in the lattice studies of the three-gluon vertex is denoted by $R$ ($\widehat{R}$ in the $B^3$ case), and will be referred to as the ``$R$-projector'' [for its exact definition, see~\1eq{Rdefinition}]; 
it receives contributions from the various form factors of the three-gluon vertex, 
both ``longitudinal'' and ``transverse'' (see Appendix~\ref{appA}). Due to the QED-like Ward identity (WI) satisfied by the $B^3$ vertex, as opposed to the Slavnov-Taylor identity (STI) satisfied by the $Q^3$ vertex,  
the former can be expressed {\it exclusively} in terms of  the $B^2$ gluon kinetic 
term $\widehat{J}(q^2)$ [with $J(q^2)=F^2(q^2)\widehat{J}(q^2)$], 
while the latter remain undetermined (they satisfy the WI automatically). 
From the kinematic point of view,  $R$ depends on the modulo of 
two independent momenta ($q^2$ and $r^2$) and the angle $\varphi$ formed between them. 
It turns out that for the special case $\varphi=\pi/2$ and $r^2=0$, 
any contribution from the ``transverse'' form factors of the three-gluon vertex drops out, 
and one finds that, quite remarkably,  $\widehat{R}(q^2)= [q^2 \widehat{J} (q^2)]^{\prime}$.

It is now relatively straightforward to recognize that 
$\widehat{J} (q^2)$ and ${J} (q^2)$  display the same type 
of logarithmic divergence in the IR, due to the masslessness of the corresponding ghost loop.
Thus, even though the precise prefactors between each ghost loop do not coincide, due to the 
difference in the form of the ghost-gluon vertex in the linear covariant ($R_{\xi}$) gauges and in the BFM   
(see discussion in Section~\ref{fullnon}), the qualitative behavior in the limit of interest is common.
Furthermore, one may recover $R(q^2)$ from  $\widehat{R}(q^2)$ by assigning tree-level values
to the ghost dressing function $F$, and the gluon-ghost kernel $H$~[see \1eq{Hdecomposition}].
Therefore, if we use the fact
that $H$ does not introduce additional IR divergences [see discussion after \1eq{darelation}], 
then $R(q^2)$ may be qualitatively modeled by  
\be
R(q^2)\sim[q^2 J (q^2)]^{\prime}. 
\label{Rmod}
\ee 

If we now employ the toy model of the previous subsection
to evaluate the rhs of~\1eq{Rmod}, 
then $R(q^2)$ is given precisely
by the expression obtained in \1eq{der4} and \1eq{der3}. Specifically, in either case
\be
R(q^2)\underset{q^2\to0}{\sim}c_2J_{a_3}(q^2),
\ee
which gives a negative logarithmic divergence in $d=4$ 
[$c_2>0$, but $\ln\left(q^2/\mu^2\right)<0$],
and a negative linear divergence in $d=3$ ($1/q >0$, but $c_2<0$). 
Summarizing, 
\be
R(q^2) \underset{q^2\to0} {\sim}\left\{
\begin{array}{ll}
\ln\left(q^2/\mu^2\right), & \quad d=4; \\
-1/q, & \quad d=3.
\end{array}
\right.
\label{IR-be}
\ee

Note that the value $q_\s{\Delta}$, which determines  the location of the maximum of $\Delta(q^2)$
corresponds now 
precisely to the ``crossing point'', $q_\s0$, namely the point where $R$ passes from positive to negative values.
Thus, {\it within this toy model}, the location of the maximum of the gluon propagator {\it coincides}
with the crossing point\footnote{If the conditions for having a maximum in the gluon propagator 
were not fulfilled, $R$ would still diverge at the origin, 
but there would be no crossing point: $R$ would be negative for all values of momenta.} 
of $R(q^2)$, \ie $q_\s{\Delta}=q_\s0$. Of course, the reason for this coincidence is directly related to the fact that 
we use a constant gluon mass, $m^2$. If instead we had employed a momentum dependent mass, $m^2(q^2)$,  
the location of these special points would differ, $q_\s{\Delta} \neq q_\s0$, as will happen in the full analysis 
of the next section. 

Finally, 
the ultraviolet behavior of the form factor $R(q^2)$ is given by
\be
R(q^2)\underset{q^2\to\infty}{\sim}\left\{
\begin{array}{ll}
1+(c_1+c_2)[1+\ln\left(q^2/\mu^2\right)], & \quad d=4; \\
1, & \quad d=3.
\end{array}
\right.
\label{R-uv}
\ee
Thus, in the three dimensional case  $R(q^2)$ saturates at its tree-level value, 
while in four dimensions it increases as a positive logarithm, $R(q^2)\to+\infty$.

\subsubsection{Numerics}

\begin{figure}[!t]
\mbox{}\hspace{-.9cm}\includegraphics[scale=.975]{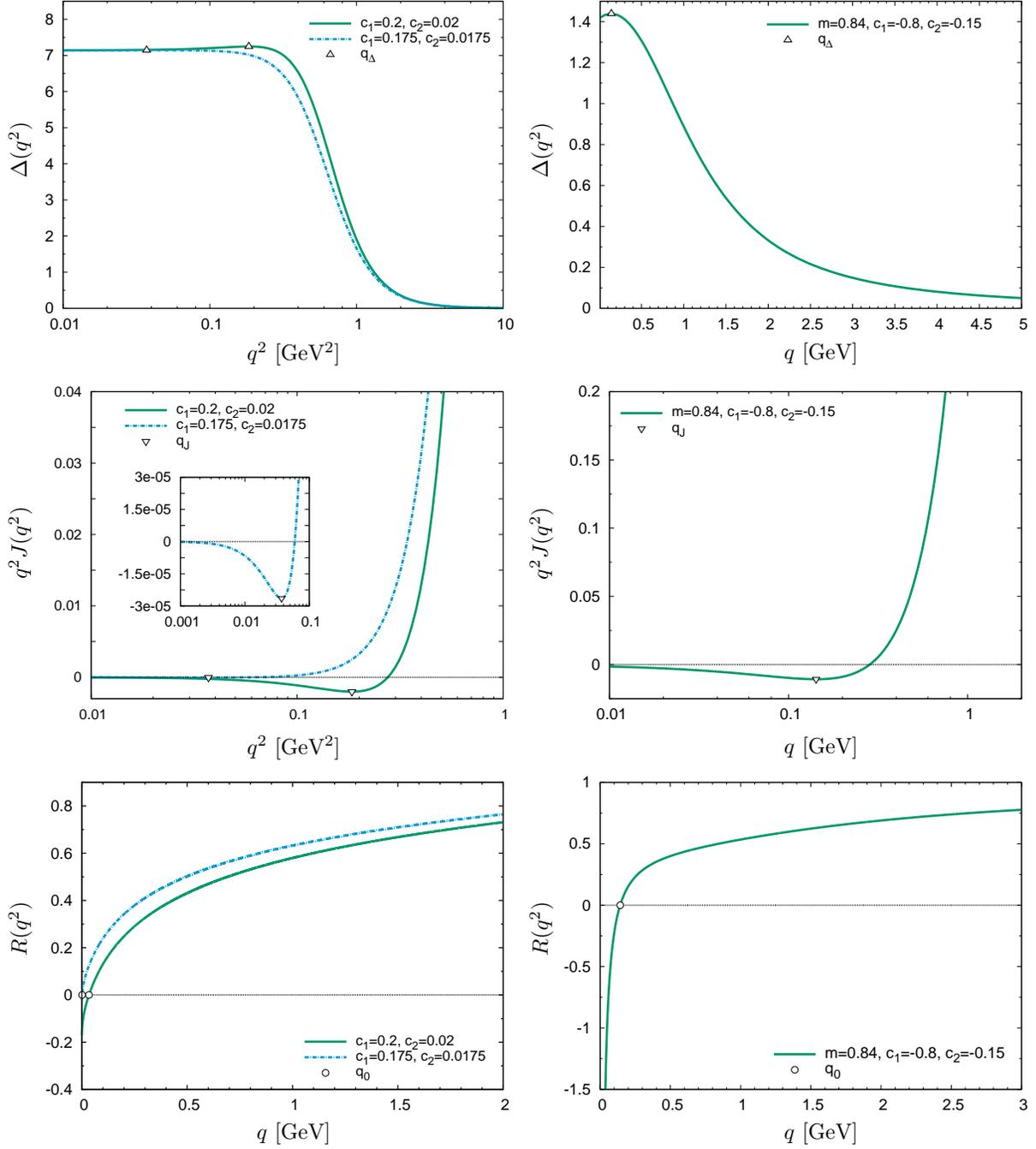}
\caption{\label{fig:toy-model}(color     online).    The    propagator
$\Delta(q^2)$, its kinetic part $q^2J(q^2)$, and the quantity $R(q^2)$
calculated in $d=3,4$ for various
values  of the  constants  $c_1$ and  $c_2$.  In $d=4$  we have 
$\mu=4.3$  GeV, \mbox{$m^2=0.14$}~GeV$^2$,  whereas in  $d=3$  we chose
$m=0.84$ GeV. Open up/down triangles and circles  mark the position of the $q_\s{\Delta}$, $q_\s J$ and $q_\s0$, respectively. Notice that, in this model, the location of these three points coincides.}
\end{figure}

In~\fig{fig:toy-model} we plot the propagator $\Delta(q^2)$,  its kinetic part $q^2J(q^2)$, 
and finally the quantity $R(q^2)\sim[q^2J(q^2)]'$ 
for some values of the $c_i$ constants  (in the four dimensional case we have 
additionally fixed\footnote{The choice $\mu =4.3$ GeV is lattice-motivated, in the sense that  
it corresponds to the last available point in the 
ultraviolet tail of the gluon propagator obtained from the simulation of~\cite{Bogolubsky:2009dc}; 
therefore, in the full non-perturbative treatment, one usually 
renormalizes the gluon propagator 
such that $\Delta^{-1}(\mu^2) = \mu^2$, at that particular point. 
Then, the IR saturation point acquires the value  $\Delta^{-1}(0)=m^2=0.14$ GeV$^2$. 
At the level of the toy model these choices simply 
help us maintain a close analogy with the full treatment presented in the next section.} $\mu$ at $4.3$ GeV, 
and, accordingly, $m^2=0.14$ GeV$^2$).

As one can see from the top panels of this figure (continuous curves),
the  propagator $\Delta(q^2)$  of~\1eq{tm-prop}  displays an  IR
peak.   In  $d=3$,  this  particular feature is  well-established,  both  at the level of the
lattice~\cite{Cucchieri:2003di,Cucchieri:2010xr}, as well as from various treatments 
in the  continuum (see, e.g., ~\cite{Aguilar:2010zx}). On the other hand, in  $d=4$, the  lattice
evidence for the appearance of such a peak is certainly inconclusive 
while  in the  continuum, to the best of our knowledge,  this possibility  has not even been
contemplated. Of course, we  hasten to  emphasize that  there are
regions in  the parameter space of  our toy model where the ``peak'' 
flattens out completely, and 
escapes detection due to numerical errors.

The kinetic part of the  propagator, $q^2J(q^2)=\Delta^{-1}(q^2)-m^2$, is plotted in the mid panels of~\fig{fig:toy-model}. 
Clearly, as mentioned earlier, due to the fact that $m^2$ is a constant, this quantity will display a (negative) minimum at exactly 
the same point where the peak of the propagator is located, $q_{\s \Delta}=q_{\s J}$. In the $d=4$ case, 
the inset shows with more accuracy the extremely shallow minimum that is obtained for precisely those values of the $c_i$ 
that cause the maximum of $\Delta(q^2)$ to flatten in the corresponding top panel.

Finally, in the bottom panels of~\fig{fig:toy-model} we plot $R(q^2)$. Again, the constancy of the gluon mass implies 
that  this quantity will cross zero exactly at the location of the propagator's peak, 
displaying afterwards the expected divergence as $q\to0$. Notice that when  $d=4$ the zero-crossing is 
clearly detectable even in the region of parameters 
where the peak of the propagator (or the minimum of the kinetic part) is barely visible. 

Summarizing, the fact that the ghost remains nonperturbatively massless  has far reaching consequences, 
that can be  captured and studied at the qualitative level by the simple model~\noeq{tm-prop}. 
As we will see in what follows, the conclusions reached in this section are robust, 
and will persist even in a fully nonperturbative setting.

\section{\label{fullnon}Full Nonperturbative analysis}

In this section we proceed to corroborate by means of a genuine nonperturbative 
analysis the qualitative picture derived from the toy model of the previous section. 
The material is organized in five interconnected subsections: \n{i} First, some general issues are discussed, which facilitate 
the perusal of what follows; 
\n{ii} Then, the detailed study of the ghost-loop contribution $J_c(q^2)$ follows, 
establishing its divergent behavior in the IR;
\n{iii} By employing the gluon mass equation and the lattice data for the 
gluon propagator, the full kinetic part, $q^2J(q^2)$ is determined, or, equivalently [by ``subtracting'' $q^2J_c(q^2)$]
the gluonic contribution $q^2J_g(q^2)$; 
\n{iv}
The $R$-projector of the three-gluon vertex  is then studied in the relevant kinematic limit, 
revealing the announced divergence;
\n{v} Finally, a detailed numerical analysis is carried out.

\subsection{Supplementary considerations}

Before entering into the technical parts of this section, 
let us briefly go over certain conceptual subtleties related to the  PT-BFM  
and its propagators~\cite{Aguilar:2006gr,Binosi:2008qk}.

Within the conventional formulation~\cite{Roberts:1994dr} of the SDE of the gluon propagator $\Delta$ ($Q^2$),  
while the {\it full} $J(q^2)$ comes out multiplied by the transverse projector $P_{\mu\nu}(q)$, 
thus reflecting the transversality of the full self-energy, no particular {\it subset} of the diagrams 
defining $J(q^2)$ displays this special property. Indeed, there is a non-trivial conspiracy of terms, stemming from each graph, that 
finally gives rise to a totally transverse self-energy. This fact is already captured at  the level of the text-book 
one-loop calculation of the gluon self-energy~\cite{Pascual:1984zb}: it is only when the ghost-loop 
is added that one obtains the required transversality.
Within this framework, therefore, the separation of ``gluonic'' and ``ghost'' contributions to $J(q^2)$ (say ``$J_g$'' and ``$J_c$'')
using Feynman graphs as a criterion is ambiguous. 

On the other hand, within the PT-BFM formalism, the SDE for the $QB$ gluon propagator $\widetilde\Delta$ possesses 
special transversality properties 
[see the discussion following \1eq{indiv-conv}, and the caption of~\fig{QB-SDE}]. As a result, 
one can meaningfully distinguish between two kinds of {\it individually transverse} 
contributions to $\widetilde{J}(q^2)$; one stemming from the ghost graphs $(a_3) + (a_4)$, to be 
denoted by ${\widetilde J}_c(q^2)$, and the rest, stemming from gluonic graphs, to be denoted by ${\widetilde J}_g(q^2)$. Thus, 
\bea
q^2 {\widetilde J}_g(q^2) P_{\mu\nu}(q) &=& [(a_1) + (a_2)]_{\mu\nu} + [(a_5) + (a_6)]_{\mu\nu},
\nonumber\\
q^2 {\widetilde J}_c(q^2)P_{\mu\nu}(q)  &=& [(a_3) + (a_4)]_{\mu\nu}.  
\label{JcJg}
\eea
Evidently, 
\be
{\widetilde J}(q^2) = 1 +  {\widetilde J}_g(q^2) + {\widetilde J}_c(q^2),
\label{tildeJcomp}
\ee
where the ``1'' on the rhs comes from the tree-level graph. 
As far as $\widetilde{J}_g(q^2)$ is concerned, it is natural to expect that it will be 
IR-finite, since it originates  from the gluonic graphs shown in~\fig{QB-SDE}, 
namely (single and double) integrals containing fully-dressed (and IR-finite) gluon propagators. 

At this point, one may use the fundamental relation of \1eq{indiv-conv}, 
which is valid for the {\it full} $J(q^2)$ and ${\widetilde J}(q^2)$, 
in order to {\it define} the corresponding $J_g(q^2)$ and $J_c(q^2)$, namely 
\be
J_{g,\,c}(q^2) = F(q^2) {\widetilde J}_{g,\,c}(q^2).
\ee
As before, 
\be
{J}(q^2) = 1 +  {J}_g(q^2) + {J}_c(q^2).
\label{normJcomp}
\ee

Finally, in order to establish a formal analogy with the toy model of the previous section, 
note that the correspondence with the terms 
appearing in \1eq{tm-prop} is 
$J_g \leftrightarrow c_1 J_{a_1}$  and $J_c\leftrightarrow c_2 J_{a_3}$. 

We next comment briefly on the 
renormalization procedure that we follow when dealing with the 
ultraviolet divergences of the $d=4$ case.
Specifically, 
we adopt the momentum subtraction (MOM) scheme, 
mainly because it is employed when renormalizing the lattice results that we use as 
inputs in our analysis.

Within the MOM scheme, the renormalized gluon propagator 
is required to assume its tree-level value at the subtraction point, \ie  
must satisfy the condition  
$\Delta^{-1}_\s{\rm R}(\mu^2) = \mu^2$, for $\mu^2 \gg m^2$.
The  (quantum) kinetic terms $J^\s{\rm R}_g(q^2)$ and $J^\s{\rm R}_c(q^2)$ are then  obtained from their 
unrenormalized counterparts through simple subtraction,  
\be 
J^\s{\rm R}_{g,\,c}(q^2) = J_{g,\,c}(q^2) - J_{g,\,c}(\mu^2).
\ee
Evidently, $J^\s{\rm R}_{g,\,c}(\mu^2) =0$. Thus, the full renormalized kinetic term is given by  
\be  
J^\s{\rm R}(q^2) = 1 + J^\s{\rm R}_{g}(q^2) + J^\s{\rm R}_{c}(q^2),
\ee
with $J^\s{\rm R}(\mu^2) =1$, consistent with the condition for 
$\Delta^{-1}_\s{\rm R}(\mu^2)$ mentioned above.

Finally, in order to keep this section as self-contained as possible, 
we list explicitly the three special values of the momentum $q$, first introduced in the context of the toy model, namely:\vspace{-.35cm}

\begin{itemize}

\item[\n{i}]  The location of the maximum of the gluon propagator, $\Delta(q^2)$, is denoted by $q_{\s \Delta}$;\vspace{-.35cm}

\item[\n{ii}]  The location of the minimum of the kinetic term, $q^2 J(q^2)$, is denoted by $q_{\s J}$;\vspace{-.35cm}

\item[\n{iii}]  The location of the zero-crossing of the $R$-projector is denoted by $q_{\s 0}$. 

\end{itemize}

\subsection{Nonperturbative ghost-loops and the minimum of the kinetic term}

Our starting point are the two fully-dressed diagrams  $(a_3)$ and $(a_4)$ of~\fig{QB-SDE}, which, according 
to the above discussion,  define $J_c(q^2)$ through 
\be
q^2 J_c(q^2) P_{\mu\nu}(q) = F(q^2)[(a_3) + (a_4)]_{\mu\nu}.
\label{Jcnp}
\ee
Since the ghost remains massless nonperturbatively, 
the resulting contribution will be IR divergent, as happens in the one-loop perturbative case.

To see this in detail, let us focus on the rhs of \1eq{Jcnp}.
Factoring out the trivial color structure $\delta^{ab}$, one has 
\begin{align}
(a_3)_{\mu\nu}&= -g^2C_A
\int_k\!(k+q)_\mu D(k)D(k+q)\widetilde{\Gamma}_{\nu}(k+q,-q,-k), 
\nonumber \\
(a_4)_{\mu\nu}&= g^2C_A g_{\mu\nu}\int_k\!D(k).
\label{gh-1ldr}
\end{align}
In the above equations,
$\widetilde{\Gamma}_{\mu}(r,q,p)$ is the PT-BFM vertex 
describing the interaction of a background gluon with a ghost and an antighost; 
unlike the conventional ghost-gluon vertex, its   
tree-level expression is symmetric in the ghost momenta, 
$\widetilde{\Gamma}^{(0)}_{\mu}(r,q,p)=(r-p)_\mu$.
In addition, we have introduced the $d$-dimensional 
measure $\int_{k}\equiv \mu^{\epsilon}/(2\pi)^{d}\!\int\!\mathrm{d}^d k$, where $\mu$ is the 't Hooft mass and $\epsilon=4-d$.

Since, due to the PT-BFM properties,  $\widetilde{\Gamma}_{\mu}$ satisfies the WI~\cite{Aguilar:2006gr,Binosi:2008qk}
\be
q^\mu\widetilde{\Gamma}_\mu(r,q,p)=D^{-1}(p)-D^{-1}(r),
\label{WIGG}
\ee
one may establish immediately the transversality of this subset of diagrams, as anticipated by the presence of the 
projector $P_{\mu\nu}(q)$ on the lhs of \1eq{Jcnp}.

One may now   
introduce an Ansatz for $\widetilde{\Gamma}_\mu$, that satisfies automatically the above WI, namely
\be
\widetilde{\Gamma}_{\mu}(r,q,p)=\frac{(r-p)_\mu}{r^2-p^2}\left[D^{-1}(r^2)-D^{-1}(p^2)\right].
\label{v-ansatz}
\ee
A corresponding construction for the conventional ghost-gluon vertex 
would be less forthcoming, given the type of STI that the latter satisfies~\cite{Pascual:1984zb}.

Obviously this procedure 
leaves the transverse part of the vertex, ${\cal A}(r,p)\left[(r\cd q)p_\mu-(p\cd q)r_\mu\right]$, 
undetermined; 
however, under rather mild assumptions on behavior of ${\cal A}(r,p)$,  
this term is subleading in the IR (see Appendix~\ref{appB}), and its effects may be neglected at this stage\footnote{Of course, in $d=4$ the omission of the transverse term 
affects the ultraviolet properties of the resulting SDE, forcing subtractive 
instead of multiplicative renormalization~\cite{Curtis:1990zs,Kizilersu:2009kg}.}.

Substituting the vertex~\noeq{v-ansatz} into the first equation of~\noeq{gh-1ldr},
and taking the trace of both sides, one obtains
\be
q^2J_c(q^2)= C_d \, F(q^2)\left[4T(q^2)+q^2S(q^2)\right],
\label{q2Jc}
\ee
where we have defined 
\be
C_d= \frac{g^2C_A}{2(d-1)},  
\ee
and
\begin{align}
T(q^2)&=
\int_k \frac{F(k+q)-F(k)}{(k+q)^2-k^2}+\left(\frac{d}2-1\right)\int_k\frac{F(k)}{k^2},\nonumber \\
S(q^2)&=\int_k\frac{F(k)}{k^2(k+q)^2}-\int_k\!\frac{F(k+q)-F(k)}{k^2[(k+q)^2-k^2]}. 
\label{SandT-final}
\end{align}

Let us now study at the IR behavior of these quantities. For the term $S$ one finds 
\begin{align}
S(q^2) &\underset{q^2\to0}{\to}\int_k \frac{F(k)}{k^4} - \int_k \frac{1}{k^2}\frac{\partial F(k)}{\partial k^2} + {\cal O}(q^2)\nonumber \\
&\sim \int_k \frac{F(k)}{k^4}- \int_0^\infty\!{\mathrm d}y\, y^{d/2-2}F'(y) + {\cal O}(q^2),
\label{infraredS}
\end{align}
where in the last step we have passed to spherical coordinates, with $y=k^2$ [see~\1eq{d-measure}]. 

The first integral on the rhs contains the divergence discussed in the 
previous section. This can be seen by simply setting  $F(y) = 1$, in which case one finds a logarithmic ($d=4$) 
or a linear ($d=3$) divergence.  Given that the full $F(y)$ saturates at a constant value in the IR, 
its presence does not qualitatively modify the behavior of the integral  
with respect to the case when $F(y)=1$; roughly speaking, it simply 
changes its prefactor from 1 to $F(0)$.  

It is now relatively 
straightforward to establish that 
the second integral in \1eq{infraredS} is subleading compared to the first one, 
as a result of the fact that  $F(y)$ is a finite function in the entire range of momenta.
Indeed, in $d=4$, integration by parts shows that it is simply equal to $F(0)$, and, therefore, it contributes a finite constant.
In $d=3$, let us assume that $F'\sim y^{-a}$; then one may naturally distinguish three cases, 
depending on the value of the exponent $a$.
\n{i} if $a<1/2$, it is clear that the integral is finite, and evidently subleading; 
\n{ii} if $1/2 \leq a<1$,  the integral diverges, but with a degree of divergence inferior to 
$1/q$ (or $y^{-1/2}$);
\n{iii} if $a \geq 1$, the second integral diverges faster than the first.
Now, given that from $F'\sim y^{-a}$ one deduces that 
$F(y) \sim  (1-a)^{-1} y^{1-a} +C$, if $a\neq 1$, and $F(y) \sim C + \ln y$, if  $a=1$, 
the finiteness of $F(y)$ imposes  the restriction $a<1$. 
Therefore, one is driven to the cases \n{i} or \n{ii}, and, consequently, 
the second integral may be finite or divergent in the IR, but is certainly subleading compared to the first.

Next, consider the term $T(q^2)$; following a similar procedure, we obtain
\be
T(q^2)\underset{q^2\to0}{\to}\ T(0) =
\int_k\!\frac{\partial F(k^2)}{\partial k^2}+\left(\frac d2-1\right)\int_k\!\frac{F(k)}{k^2}.
\label{T0}
\ee
It is now  immediate to recognize that $T(0)$ vanishes, since~\1eq{T0} is a particular case of the so-called 
``seagull-identity''~\cite{Aguilar:2009ke},  
\be
\int_k\! k^2\frac{\partial{f}(k^2)}{\partial k^2}+\frac d2\int_k\!f(k^2)=0, 
\label{seagull}
\ee 
valid in dimensional regularization\footnote{The origin of \1eq{seagull} is simple integration by parts, 
where the surface term is dropped by appealing to 
the analyticity properties of dimensional regularization. Its main function  
in the context of gluon mass generation is to enforce 
the complete cancellation of all quadratic (seagull-type) divergences.}.

Then, employing that $T(0)=0$, and after repeated use of~\noeq{seagull}, one finds that
\be
T(q^2) \underset{q^2\to0}{\to}\ -\frac{1}{12}\left(d-2\right)q^2\!\int_k \frac{1}{k^2}\frac{\partial F(k)}{\partial k^2} + {\cal O}(q^4),
\label{TatOq2}
\ee
that is, we end up with the first integral on the rhs of \1eq{infraredS}, which is subleading.

Thus, if we split $J_c$ in a part that contains the leading contribution in the IR, $J_c^\ell(q^2)$, 
and the rest that is subleading, $J_c^{\mathrm{s}\ell}(q^2)$,    
\be
J_c(q^2) = J_c^\ell(q^2) + J_c^{\mathrm{s}\ell}(q^2), 
\label{Jcsplit}
\ee
we conclude that the leading divergent contribution  
is that contained in the first term of~$S(q^2)$, namely
\be
J_c^\ell(q^2) = C_d F(q^2) \int_k\frac{F(k)}{k^2(k+q)^2}.
\label{Jcdiv}
\ee 
On the other hand, $J_c^{\mathrm{s}\ell}(q^2)$ consists of all those 
terms that have been discarded throughout the procedure described above. 
It may be computed numerically, but its detailed form is of no immediate interest, and it 
will be simply included in the full curve describing $J_c(q^2)$.

The divergent nature of  $J_c(q^2)$ causes the kinetic term $q^2J(q^2)$
to acquire a minimum in the IR region, as can be demonstrated by 
following basically the arguments related with \1eq{der4}.
In particular, using \2eqs{normJcomp}{Jcsplit}, one has  
\be
[q^2J(q^2)]' = J_c^\ell(q^2) +  
\left\{ 1+ q^2 J^{\,\prime}(q^2) + J_c^{\mathrm{s}\ell}(q^2)\right\}.
\label{kinprime}
\ee
Now, as happens in the case of the toy model, 
\n{i} the quantity in curly brackets is subleading in the IR, and \n{ii}
the above derivative is positive in the ultraviolet, since $q^2J(q^2)$ increases 
[and so, $\Delta(q^2)$ decreases].
Thus, the derivative reverses its sign, becoming zero at the point $q_{\s J}$, namely 
\be
[q^2J(q^2)]'_{q=q_{\s J}} =0.
\ee
We must emphasize at this point that even though the existence of the minimum is established 
by means of the above argument, its location cannot be accurately determined, because we do not know  
all terms appearing in the curly bracket of \1eq{kinprime}. Therefore, $q_{\s J}$ {\it cannot} be 
computed  {\it directly}; however, in the next subsection we will determine its value  {\it indirectly}, 
from the (better known) combination $\Delta^{-1}(q^2)-m^2(q^2)$.  

\subsection{Maximum of $\Delta(q^2)$, and indirect determination of $J_g(q^2)$.}

Let us now examine whether the maximum 
of the gluon propagator established in Section~\ref{toy} in the context of the toy model 
persists in the full nonperturbative treatment. 

Evidently, the main qualitative difference between the two situations 
is that now the gluon mass is not a constant, but a function of the momentum, $m^2(q^2)$.
Specifically, the corresponding $[\Delta^{-1}(q^2)]^{\prime}$ reads 
\be
[\Delta^{-1}(q^2)]^{\prime} = [q^2J(q^2)]' + [m^2(q^2)]^{\prime}.
\label{dernonp}
\ee
Now, if the quantity  $[m^2(q^2)]^{\prime}$ is a ``well-behaved'' function, 
then the arguments following  \1eq{kinprime} would again go through here.
In particular, the existence of a zero is guaranteed,  
provided that the IR divergence of $J_c^\ell(q^2)$ is not cancelled exactly by a 
similar divergence (with opposite sign) contained in $[m^2(q^2)]^{\prime}$.

To discard this remote possibility, we turn to the 
dynamical equation that governs $m^2(q^2)$, which, in its exact form, reads~\cite{Binosi:2012sj}  
\be     
m^2(q^2)    =     
-g^2 C_AD(q^2)
\int_k m^2(k^2) \Delta_\rho^\mu(k)\Delta^{\nu\rho}(k+q)
{\cal K}_{\mu\nu}(k,q),
\label{masseq}
\ee
with the kernel given by
\bea
{\cal K}_{\mu\nu}(k,q) &=& [(k+q)^2 - k^2] \left\{ 1 - [\Y(k+q) + \Y(k)]\right\}g_{\mu\nu}
\nonumber \\
&+& [\Y(k+q)-\Y(k)](q^2 g_{\mu\nu}-2q_\mu q_\nu),
\label{massK}
\eea
and $\Y$ defined through 
\be
\Y(k^2)=\frac{g^2 C_A}{4 k^2} \,k_\alpha\! \int_r\!\Delta^{\alpha\rho}(r)
\Delta^{\beta\sigma}(r+k)\Gamma_{\sigma\rho\beta}(-r-k,r,k),
\label{theY}
\ee
where $\Gamma_{\sigma\rho\beta}$ is the full three-gluon vertex.  
 
This equation has been studied in $d=4$, under certain simplifying assumptions 
regarding the structure of the function $\Y(k^2)$. 
In particular, $\Y$ was replaced by its  
lowest perturbative approximation, given by
\be
Y_{\mathrm R}(k^2)=-\frac {\alpha_s C_A}{4\pi}\frac{15}{16}\log\frac{k^2}{\mu^2},
\label{Yappr}
\ee
with $\alpha_s$ the value of the strong coupling at the subtraction point chosen. 
In practice, this simple approximation is improved by letting $\Y\to C\Y$, where $C$ 
is an arbitrary constant,  modeling further  corrections that may  be added  to the ``skeleton'' result of~\1eq{Yappr}.

From \1eq{masseq} one 
obtains positive-definite solutions for the gluon mass function, at least within a reasonable range 
of physical momenta. 
In particular, for $\alpha_s=0.22$, which is the ``canonical'' MOM value for \mbox{$\mu=4.3$ GeV}~\cite{Boucaud:2005rm}, 
and $C=9.2$, the function  
$m^2(q^2)$ is positive and monotonically decreasing in the range of momenta  between\footnote{Past this point $m^2(q^2)$ 
turns negative (but its magnitude is extremely small), 
reaching finally zero from negative values~\cite{Holdom:2013uya}. The necessary refinements for 
rectifying this will be reported elsewhere.} 0 to \mbox{$5.5$ GeV}.

\begin{figure}[!t]
\mbox{}\hspace{-0.6cm}\includegraphics[scale=1]{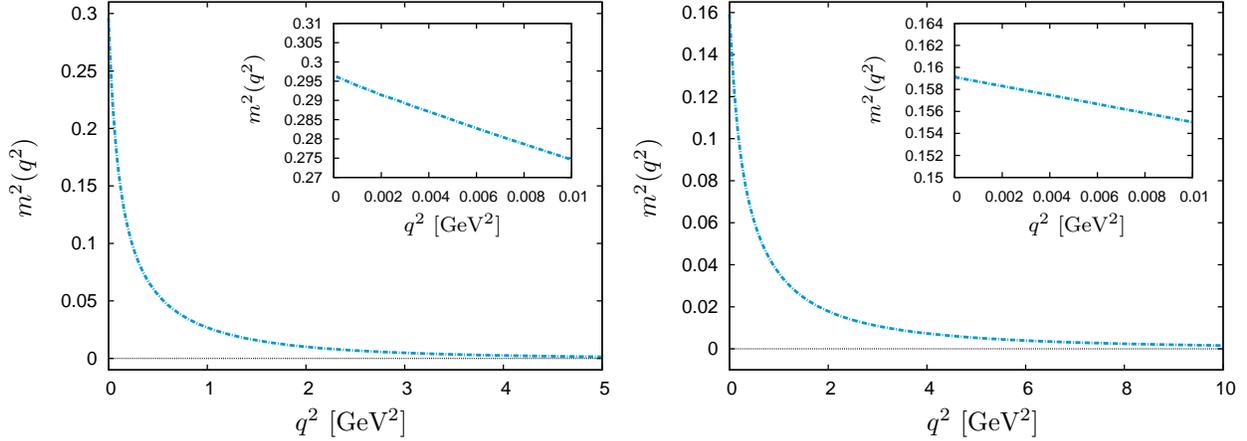}
\caption{\label{fig:mass-4d}(color online). Solutions of the mass equation for the SU(2) (left) and SU(3) (right) gauge groups. In the insets we show the deep IR region of the same curves.}
\end{figure}

Evidently, this behavior (see~\fig{fig:mass-4d})
excludes the possibility of $m^2(q^2)$ having a divergent (and positive) derivative at the origin. 
Therefore, we conclude that the rhs of \1eq{dernonp} must reverse its sign at some point ($q_{\s \Delta}$), 
where the corresponding gluon propagator will display a maximum. 
Of course, the presence of the term $[m^2(q^2)]^{\prime}$ in \1eq{dernonp} prevents the 
coincidence between $q_{\s \Delta}$ and $q_{\s J}$, unless 
$[m^2(q^2)]^{\prime}_{q=q_{\s J}} =0$;  but this possibility is discarded, 
due to the monotonic nature of $m^2(q^2)$. Thus, we conclude that $q_{\s \Delta}\neq q_{\s J}$. 

One may go one step further, and demonstrate that, in fact,  $q_{\s J} < q_{\s \Delta}$.
Indeed, evaluating both sides of \1eq{dernonp} at the point $q_{\s J}$, where the first term on the  rhs vanishes,
one has 
\be
[\Delta^{-1}(q_{\s J}^2)]^{\prime} = [m^2(q_{\s J}^2)]^{\prime} < 0,
\label{dernonp1}
\ee
since $m^2(q^2)$ is monotonically decreasing. Thus, at $q_{\s J}$ the derivative 
of $\Delta^{-1}(q^2)$ is still negative, and has yet to reach its  
point of zero-crossing, a fact that places  $q_{\s \Delta}$ to the right of $q_{\s J}$ on the axis of momenta. 
As shown in Fig.~\ref{fig:qDeltavsqJ},
this particular inequality is in complete agreement with the numerical analysis presented in the last subsection.

In addition to consolidating the above considerations,
the (approximate) knowledge of $m^2(q^2)$, when combined with the lattice information 
on the full $\Delta(q^2)$~\cite{Cucchieri:2007md,Cucchieri:2010xr,Bogolubsky:2009dc,Oliveira:2009eh,Ayala:2012pb},
may furnish the full $q^2J(q^2)$, simply from 
\be
q^2J(q^2)= \underbrace{\Delta^{-1}(q^2)}_{\mathrm{lattice}}-\underbrace{m^2(q^2)}_{\mathrm{\1eq{masseq}}}.
\label{ind-kinetic}
\ee 

When coupled with the result~\noeq{q2Jc}, the procedure outlined above allows for a 
gauge-invariant identification of the two components contributing to the gluon propagator. 
Specifically, the gluon kinetic term can be obtained from [see also \1eq{normJcomp}]
\be
q^2+q^2J_g(q^2)=\underbrace{\Delta^{-1}(q^2)}_{\mathrm{lattice}}-\underbrace{m^2(q^2)}_{\mathrm{\1eq{masseq}}}-\underbrace{q^2J_c(q^2)}_{\mathrm{\1eq{q2Jc}}}.
\label{approxkin}
\ee
The results of these operations will be discussed in the last subsection. 

\subsection{The $R$-projector}

The final step is to link the gluon kinetic term with 
lattice simulations of the three-gluon vertex.
To this end, let us recall that the typical quantity employed on
the lattice projects the full vertex on its tree-level value,
dividing out, at the same time, external leg corrections~\cite{Cucchieri:2006tf,Cucchieri:2008qm}.
Specifically, for the three-gluon vertex in the Landau gauge, one considers 
\be
R(q,r,p) = \frac{{\cal N}(q,r,p)}{{\cal D}(q,r,p)},
\label{Rdefinition}
\ee
where 
\begin{align}
 {\cal N}(q,r,p) &= \Gamma^{(0)}_{\alpha\mu\nu}(q,r,p)P^{\alpha\rho}(q)P^{\mu\sigma}(r)P^{\nu\tau}(p)\Gamma_{\rho\sigma\tau}(q,r,p), \nonumber \\
{\cal D}(q,r,p) &=\Gamma^{(0)}_{\alpha\mu\nu}(q,r,p)P^{\alpha\rho}(q)P^{\mu\sigma}(r)P^{\nu\tau}(p)\Gamma^{(0)}_{\rho\sigma\tau}(q,r,p).
 \label{NandD}
\end{align}
As already mentioned, the ratio~\noeq{Rdefinition} can be characterized by the modulo of two independent momenta and the angle formed between them, so that one has $R=R(q^2,r^2,\varphi)$. Then, the quantity of interest 
corresponds to the so-called ``orthogonal configuration'', $\varphi=\pi/2$, where, in addition, 
we take the limit $r^2\rightarrow 0$, namely  $R(q^2,0,\pi/2)$.

In this particular limit, $R$ may be obtained from the combination~\cite{Binosi:2013rba}
\be
R(q^2,0,\pi/2) = X_7(q^2,0,\pi/2) + q^2 X_9(q^2,0,\pi/2).
\label{R0}
\ee
In the formula above, $X_{7,9}$ represent two of the ten longitudinal form factors which characterize the longitudinal part of the vertex (see~Appendix~\ref{appA}); their explicit form can be determined by solving the STI satisfied by the vertex and reads~\cite{Ball:1980ax}
\begin{align}
X_7(q,r,p) &= \frac{1}{4}\lbrace 2[F(q^2)J(p^2)a_{rqp} + F(p^2)J(q^2)a_{rpq}] + r^2[F(p^2)J(r^2)b_{qpr} + F(q^2)J(r^2)b_{pqr}] \nonumber \\
&+ (q^2-p^2)[F(r^2)J(q^2)b_{prq}+F(q^2)J(p^2)b_{rqp}-F(r^2)J(p^2)b_{qrp}-F(p^2)J(q^2)b_{rpq}] \nonumber \\
&+ 2(q\spr r)F(p^2)J(q^2)d_{rpq}+2(r\spr p)F(q^2)J(p^2)d_{rqp}\rbrace,\label{X7general}\\
X_9(q,r,p) &= \frac{F(r^2)}{q^2-p^2}[J(q^2)a_{prq}-J(p^2)a_{qrp}+(r\spr p)J(p^2)d_{qrp}-(q\spr r)J(q^2)d_{prq}],
\label{X9general}
\end{align}
with $a$, $b$ and $d$ representing the form factors appearing in the tensorial decomposition of the gluon-ghost kernel $H$~\cite{Ball:1980ax}
\be
H_{\nu\mu}(p,r,q) = g_{\mu\nu}a_{qrp} - r_\mu q_\nu b_{qrp} + q_\mu p_\nu c_{qrp} + q_\nu p_\mu d_{qrp} + p_\mu p_\nu e_{qrp},
\label{Hdecomposition}
\ee
and $a_{qrp}$ a short-hand notation for $a(q,r,p)$, etc.
Note finally that, in this particular kinematic limit,
the four (undetermined) transverse components of the three-gluon vertex (see~Appendix~\ref{appA})  
drop out completely. 

Let us now study the IR behavior of $R(q^2,0,\pi/2)$. 
Consider first the $X_7$ term; in the orthogonal configuration one has
\be
p^2 = q^2+r^2; \quad (q\spr r)=0; \quad (q\spr p)=-q^2; \quad (r\spr p)=-r^2; 
\label{orthogonalconf}
\ee
then,
taking the limit $r^2\rightarrow 0$,  \1eq{X7general} gives the result
\be
X_7(q^2,0,\pi/2) = F(q^2)J(q^2)a(0,q,-q).
\label{X7zero}
\ee

For $X_9$, which in the orthogonal configuration reads
\be
X_9(q^2,r^2,\pi/2) = F(r^2)J(p^2)d_{qrp} - \frac{F(r^2)}{r^2}[J(q^2)a_{prq}-J(p^2)a_{qrp}],
\label{X9orthogonal}
\ee
the corresponding treatment is slightly more involved. 
The first term appearing in \1eq{X9orthogonal} can be simplified using the identity\footnote{This identity is a direct consequence of the STI satisfied by the ghost kernel $H$, and constitutes a necessary condition for obtaining a consistent solution of the STIs of the three-gluon vertex~\cite{Binosi:2011wi}.}~\cite{Binosi:2011wi} 
\be
F(r^2)[a_{qrp}-(q\spr r)b_{qrp}+(q\spr p)d_{qrp}]=F(q^2)[a_{rqp}-(q\spr r)b_{rqp}+(p\spr r)d_{rqp}],
\label{STIconstraint}
\ee
yielding in the limit of interest
\be
q^2 F(0)d(q,0,-q) = F(0) - F(q^2)a(0,q,-q).
\label{darelation}
\ee
In arriving at the above result 
we used the fact that, in the Landau gauge, $a(q,0,-q)$ 
maintains to all orders its tree-level value~\cite{Ibanez:2012zk}, \ie $a(q,0,-q)=1$.

For the second term in~\1eq{X9orthogonal} one needs to perform a Taylor expansion around $r^2=0$ of both $a_{prq}$ and $J(p)a_{qrp}$; using
\begin{align}
a_{prq} &\underset{r^2\to0}{=} 1 +\left. r^2\frac{\partial}{\partial r^2}a_{prq}\right\vert_{r^2=0}+{\cal O}(r^4),\nonumber \\
J(p)a_{qrp} &\underset{r^2\to0}{=}  J(q) + \left.r^2\frac{\partial}{\partial r^2}\left[a_{qrp}J(p)\right]\right\vert_{r^2=0}+{\cal O}(r^4),
\label{Taylorexp}
\end{align}
we obtain
\be
-\frac{F(r^2)}{r^2}\left[J(q^2)a_{prq}-J(p^2)a_{qrp}\right] \underset{r^2\to0}{=} F(0)\left[J'(q^2) +\left. J(q^2)\frac{\partial}{\partial r^2}(a_{qrp}-a_{prq})\right\vert_{r^2=0}\right],
\label{lim}
\ee
where the prime denotes, as usual, derivatives with respect to $q^2$. Thus, inserting \2eqs{darelation}{lim} in~\1eq{X9orthogonal}, one derives the expression
\be
q^2 X_9(q^2,0,\pi/2) = J(q^2)[F(0) - F(q^2)a(0,q,-q)] + F(0)q^2J'(q^2) + {\cal O}(q),
\label{X9zero}
\ee
with ${\cal O}(q)$ indicating subleading terms, specifically the derivative of $a_{qrp}-a_{prq}$ appearing in~\1eq{lim}.

Substituting the results~\noeq{X7zero} and~\noeq{X9zero} in \1eq{R0}, we obtain the final expression (we only indicate $q^2$ in the argument)
\be
R(q^2) = F(0)[q^2J(q^2)]' + R^{\mathrm{s}\ell}(q^2),
\label{Rfinal}
\ee
which shows that the behavior of $R$ in the deep IR is determined solely by $J$, 
\be
R(q^2)\underset{q^2\to0}{\sim}F(0)J(q^2).
\ee
The term $R^{\mathrm{s}\ell}(q^2)$ denotes the subleading corrections not 
contained in the first term. 
Therefore, from \1eq{Jcdiv}, the dominant contribution in that limit is  
\be
R(q^2)\underset{q^2\to0}{\sim} C_d \, F^2(0) \int_k\frac{F(k)}{k^2(k+q)^2}.
\label{Rdivfinal}
\ee
If we now assume that the ultraviolet behavior of $R(q^2)$ is qualitatively described by \1eq{R-uv},  
then the  $R(q^2)$ of \1eq{Rfinal} must vanish  at a point $q_{\s 0}$, 
$R(q_{\s 0}^2)=0$, and then eventually diverge in the IR, according to \1eq{Rdivfinal}. 

The location of $q_{\s 0}$ with respect to the other two special points, $q_{\s \Delta}$ and  $q_{\s J}$, 
is not possible to determine. Since, from  \1eq{dernonp} we have $[q_{\s \Delta}^2J(q_{\s \Delta}^2)]' = - [m^2(q_{\s \Delta}^2)]^{\prime}$, 
the value of  $R(q^2)$ at these two points is given by 
\begin{align}
R(q_{\s \Delta}^2 ) &= R^{\mathrm{s}\ell}(q_{\s \Delta}^2) - F(0) [m^2(q_{\s \Delta}^2)]^{\prime},
\nonumber\\
R(q_{\s J}^2 ) &= R^{\mathrm{s}\ell}(q_{\s J}^2).
\label{Rqstar}
\end{align}
For the  point $q_{\s 0}$ to coincide with either $q_{\s \Delta}$ or $q_{\s J}$, the corresponding rhs in \1eq{Rqstar} 
ought to vanish; this possibility, however, cannot be checked analytically, due to the lack of knowledge of the function  
$R^{\mathrm{s}\ell}$.
Of course, in the toy model,  $R^{\mathrm{s}\ell}$ is 
identically zero, and so is the derivative of the mass, forcing the equality between these three special points. 
The available lattice data for SU(2) in $d=3$~\cite{Cucchieri:2010xr,Cucchieri:2006tf,Cucchieri:2008qm} 
seem to suggest a relative proximity between  $q_{\s 0}$ and $q_{\s \Delta}$, 
with $q_{\s \Delta}\approx q_{\s 0}\approx380$ MeV; 
of course, the  lattice parameters used for computing  
the two- and three-point functions are rather different, so this comparison is only suggestive at this point.  

\subsection{Numerical results}

\begin{figure}[!t]
\mbox{}\hspace{-1.4cm}\includegraphics[scale=.975]{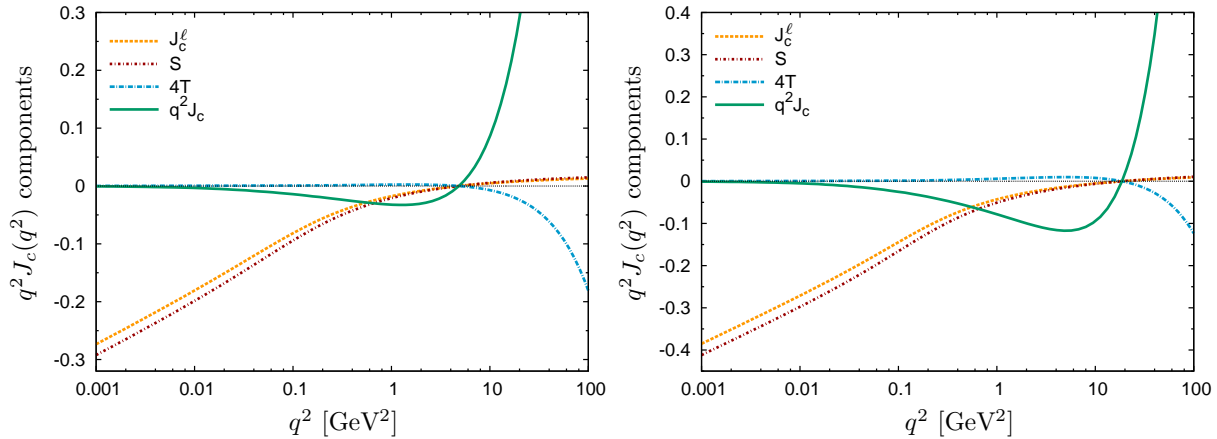}
\caption{\label{fig:q2Jc4d-4d}(color online). The ghost-loop contribution,  $q^2J_c(q^2)$, to the gluon kinetic term $q^2J(q^2)$ for the SU(2) (left) and SU(3) (right) gauge groups.}
\end{figure}

We finally carry out a detailed numerical study of all the quantities introduced in the previous four subsections.

\begin{figure}[!t]
\mbox{}\hspace{-1.4cm}\includegraphics[scale=.975]{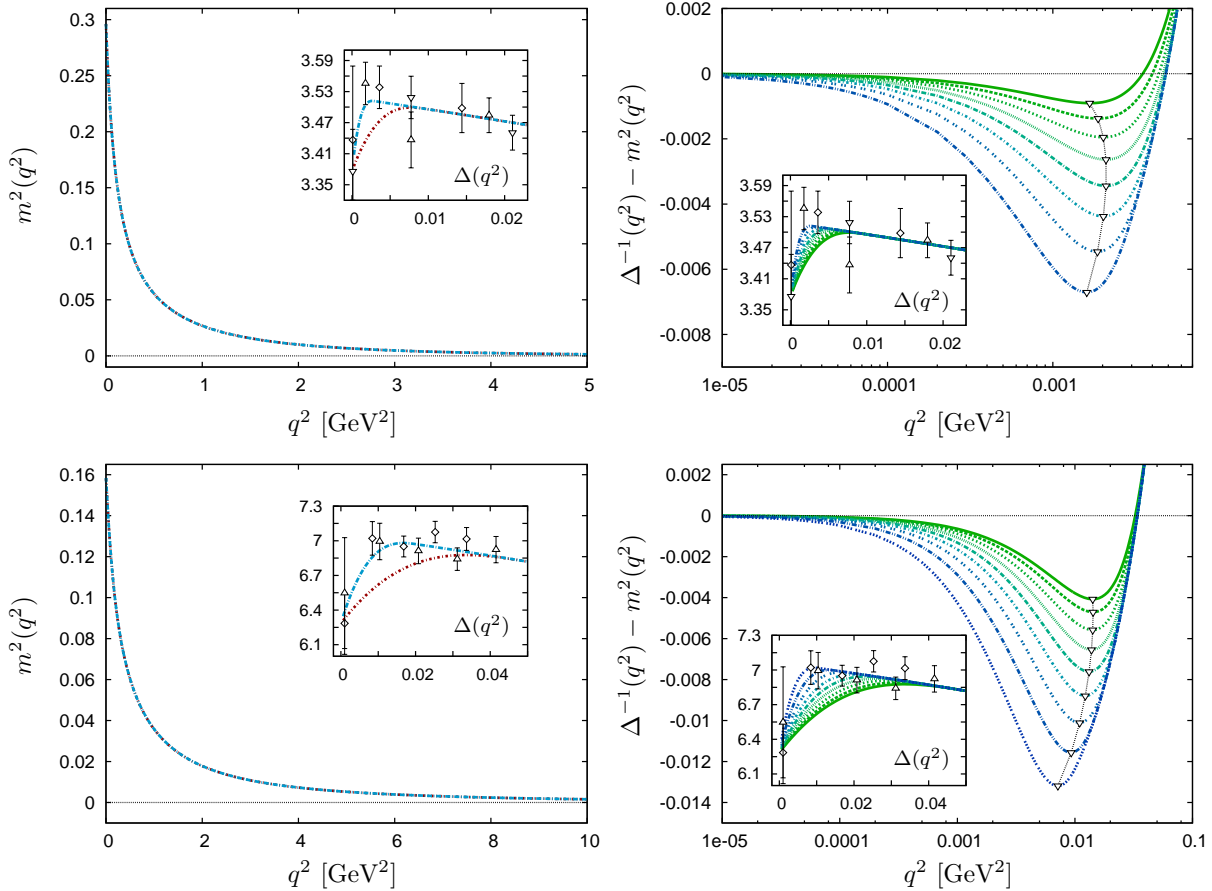}
\caption{\label{fig:mass-lake-4d}(color online). The dynamical gluon mass (left panels) and the propagator's full kinetic part $\Delta^{-1}(q^2)-m^2(q^2)$ (right panels) for the SU(2) (top) and SU(3) (bottom) gauge groups. Whereas the solutions of the mass equation are clearly insensitive to the presence of a maximum in the propagator, as shown for two representative cases, the full kinetic term develops a negative minimum ($q_\s{J}$), whose position is marked in the right panels by open (down) triangles. Insets show in all cases the IR behavior of the various propagator fits used as input, together with the corresponding lattice data.} 
\end{figure}

We start with the four dimensional case, for which in~\fig{fig:q2Jc4d-4d} we show the full non-perturbative results for the ghost-loop contribution~\noeq{Jcnp} to the gluon kinetic term~$q^2J(q^2)$. For obtaining these results we have evaluated all the terms appearing in~\2eqs{q2Jc}{SandT-final}, using as input the SU(2)~\cite{Cucchieri:2010xr} and SU(3)~\cite{Bogolubsky:2009dc} unquenched lattice data for the ghost dressing function. As anticipated, the IR logarithmic divergence, clearly identified by the linear behavior (in log scale) of the $S$ term, persists in both cases even in a fully nonperturbative setting.  

As a consequence of this divergence, the lattice data for the gluon propagator 
must have a maximum located in the (deep) IR region. It turns out that such a maximum is indeed already 
encoded in the lattice data for
the gluon propagator, which display a suppression of the deep IR points. This particular feature is 
shown in the insets appearing in the panels of~\fig{fig:mass-lake-4d}, where, together with the lattice data, 
we also plot different fitting curves, in which the position of the maximum is varied.

\begin{figure}[!t]
\mbox{}\hspace{-.8cm}\includegraphics[scale=.975]{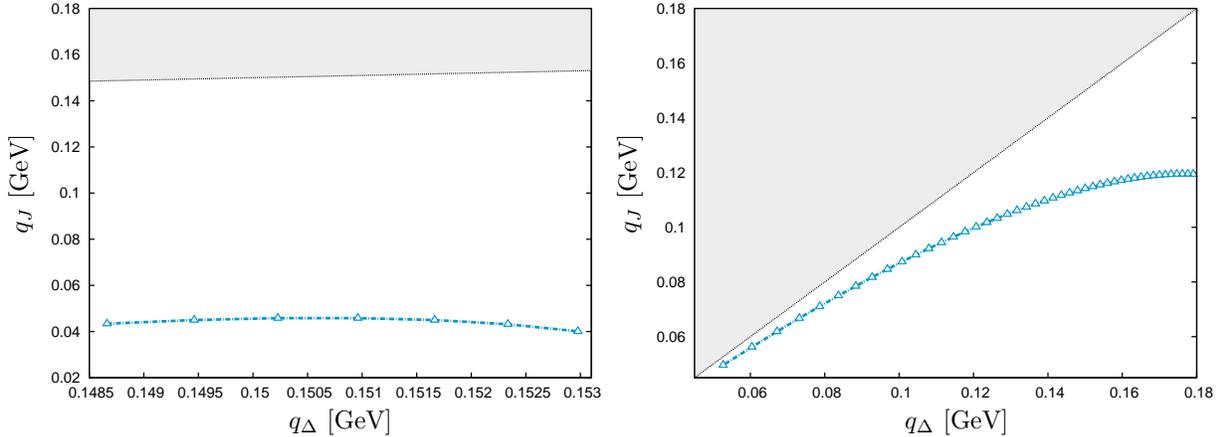}
\caption{\label{fig:qDeltavsqJ}(color online). 
The points $q_{\s J}$ plotted as a function of the points $q_{\s \Delta}$ for the SU(2) (left) and SU(3) (right) gauge groups.
The shaded area on both panels corresponds to the region $q_{\s J}\ge q_{\s \Delta}$.}
\end{figure}

\begin{figure}[!t]
\mbox{}\hspace{-1.4cm}\includegraphics[scale=.975]{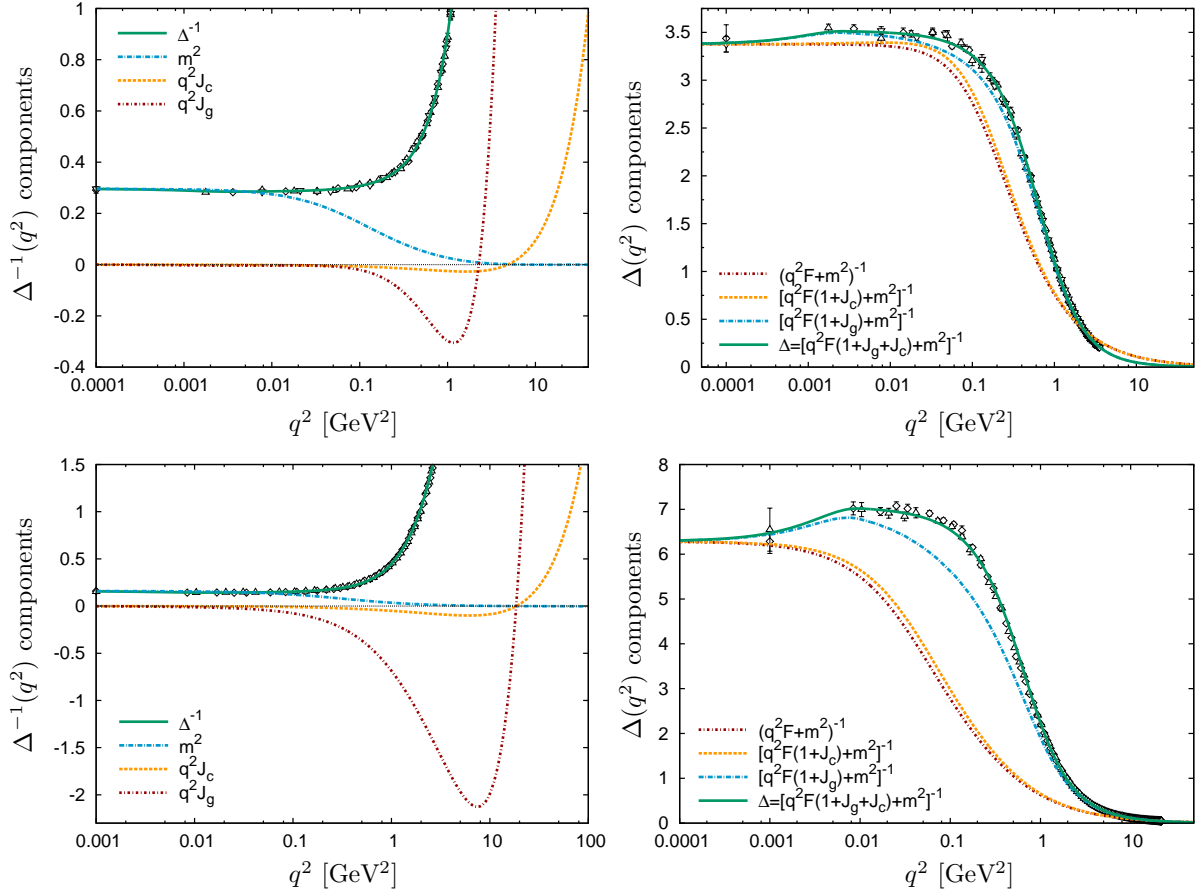}
\caption{\label{fig:gauge-invariant-4d}(color online). The gauge-invariant decomposition of the ghost and gluon components of the gluon propagator (right panels) and its inverse (left panels) for the SU(2) (top) and SU(3) (bottom) gauge groups. Notice that in the left panels the tree-level contribution to the gluon kinetic term has been suppressed; its effect can be seen clearly in Fig~\ref{fig:Jg-4d}.}
\end{figure}

\begin{figure}[!t] 
\mbox{}\hspace{-1.4cm}\includegraphics[scale=.975]{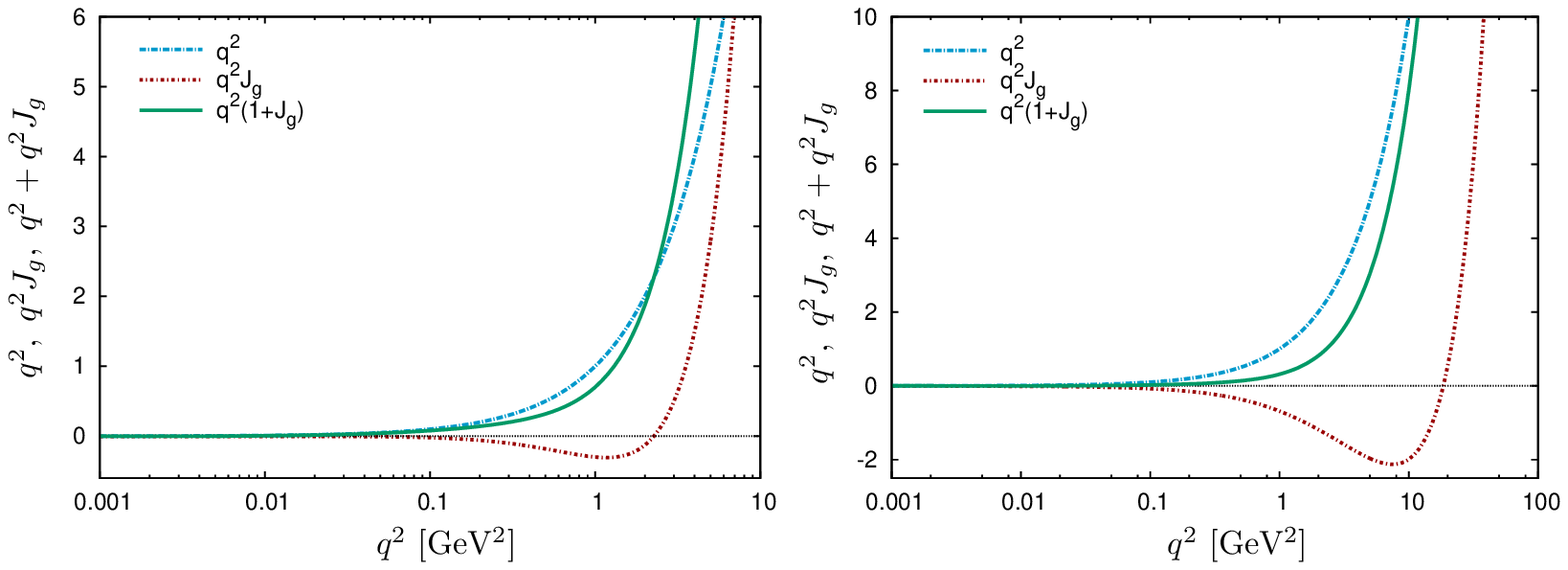}
\caption{\label{fig:Jg-4d}(color online). Tree-level and quantum contributions to the gluon kinetic term for SU(2) (left panel) and SU(3) (right panel).}
\end{figure} 

We next turn to the indirect determination of $q^2 J(q^2)$ from \1eq{ind-kinetic}, using as basic input  
the family of curves for $\Delta(q^2)$ obtained in the previous step. To that end, we first establish that,  
when the latter curves are used as input for the mass equation~\noeq{masseq}, 
the resulting masses turn out to be completely independent of the location and the size of the 
maximum of the propagator (left panels of~\fig{fig:mass-lake-4d}).
Thus, while $\Delta(q^2)$ is varied on the rhs of \1eq{ind-kinetic}, the corresponding $m^2(q^2)$ remains the same for all cases.
The result of this procedure is shown 
on the right panels of~\fig{fig:mass-lake-4d}; 
each $q^2 J(q^2)$ so obtained vanishes 
at the origin, decreases in the deep IR, 
and reaches a negative minimum before crossing zero and turning positive. The location of the 
corresponding minimum, $q_{\s J}$, is clearly marked for each separate case. 

Evidently, since every $\Delta(q^2)$ has a maximum at a point $q_{\s \Delta}$, and since from each such 
$\Delta(q^2)$ we obtain a $q^2 J(q^2)$ with a minimum at a point $q_{\s J}$, one may 
plot $q_{\s J}$ as a function of  $q_{\s \Delta}$. The resulting relation is shown in Fig.~\ref{fig:qDeltavsqJ}.
The shaded area serves as a reference, corresponding to the case  $q_{\s J}\ge q_{\s \Delta}$. 
The plot clearly indicates 
that all points lie below this region, demonstrating that, indeed,  $q_{\s J}<q_{\s \Delta}$,
 as previously anticipated using~\1eq{dernonp1}. 

Notice that, as an important by-product of this analysis, 
we are able to disentangle gauge-invariantly the ghost and gluon contributions 
to the kinetic term and the propagator. 
This is shown in~\fig{fig:gauge-invariant-4d}, where one can appreciate 
that the gluon contribution is dominant; however, in the crucial region 
below 1 GeV$^2$ there are still sizable contributions from ghost terms. 
It should be also noticed that there is a substantial difference 
between the SU(3) and SU(2) gauge groups, in the sense that the 
relative size disparity between the various components is more moderate in the SU(2) case.
Finally, in~\fig{fig:Jg-4d}  we show the tree-level and quantum part 
of the gluon kinetic term: notice that in this case, any structure appearing 
in the quantum term gets completely washed out 
by the tree-level term (obviously absent for the ghost contributions).

\begin{figure}[!t]
\includegraphics[scale=0.9]{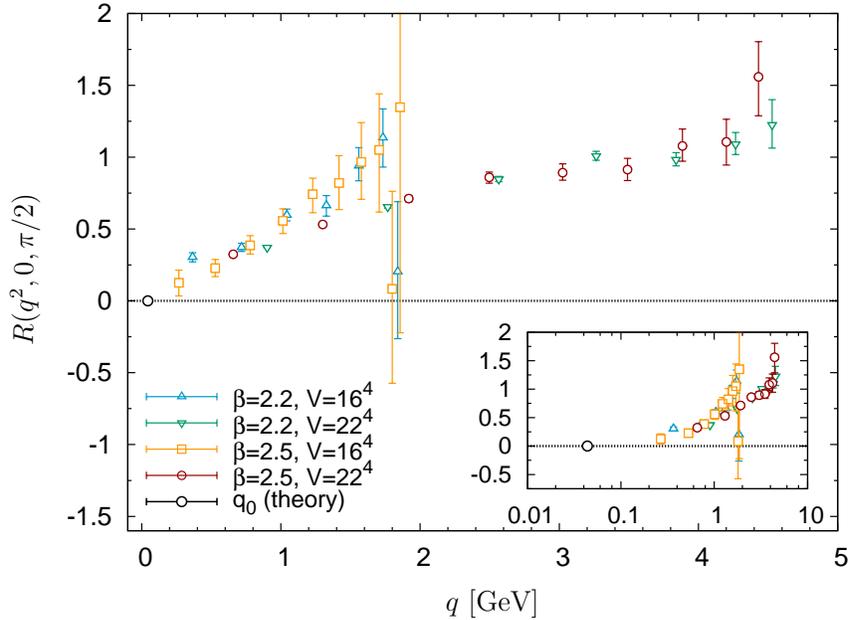} 
\caption{\label{fig:3gluon-4d}(color online). Prediction for the zero-crossing of the SU(2) form factor $R(q^2,0,\pi/2)$ measured on the lattice in four dimensions. The inset shows a logarithmic plot of the same quantity.}
\end{figure}

As~\1eq{Rfinal} reveals, the position of the minimum of the full kinetic 
term provides an estimate for the momentum $q_\s0$ where the three-gluon  
projector $R(q^2,0,\pi/2)$ crosses zero and reverses sign. In the SU(2) 
case this turns out to be located quite deep in the IR, 
as one gets (see~\fig{fig:3gluon-4d}) $q_\s0\approx44$ MeV, 
while for SU(3) we obtain $q_\s0\approx132$ MeV. 

At this point, we can use the relation (see, \eg~\cite{Rothe:1992nt})
\be
q=\frac2a\sin\frac{\pi k}{L} \times 197.3\ \mathrm{MeV}\ \mathrm{fm}, 
\ee
with $a$ the lattice size (in fermi), $L$ the number of lattice sites and $k\leq L$ an integer 
locating the different sites in the corresponding lattice direction, in order to convert 
the numbers obtained above into the lattice volumes needed to resolve them. 
Setting $k=1$ (corresponding evidently to the minimum momentum which can be 
reached for a given $L$), and choosing the most coarse lattices used in the 
literature ($a\approx 21$ fm at $\beta=2.5$ for the SU(2) case~\cite{Cucchieri:2008qm}), 
one obtains $L\sim130$, 
which does not seem attainable with current simulations (which have $L=22$ at most). 
In the SU(3) case, assuming that simulations can be performed at $\beta=5.7$ with $a\approx0.17$ fm (that is, in the same 
conditions used for simulating the gluon and ghost two-point sectors in~\cite{Bogolubsky:2009dc}), 
one obtains instead~$L\sim60$.  
Obviously, these numbers are indicative, but they seem to suggest that resolving the zero-crossing of $R$ on the lattice 
in four dimensions could represent a challenging endeavor. 

\begin{figure}[!t] 
\mbox{}\hspace{-1.4cm}\includegraphics[scale=.975]{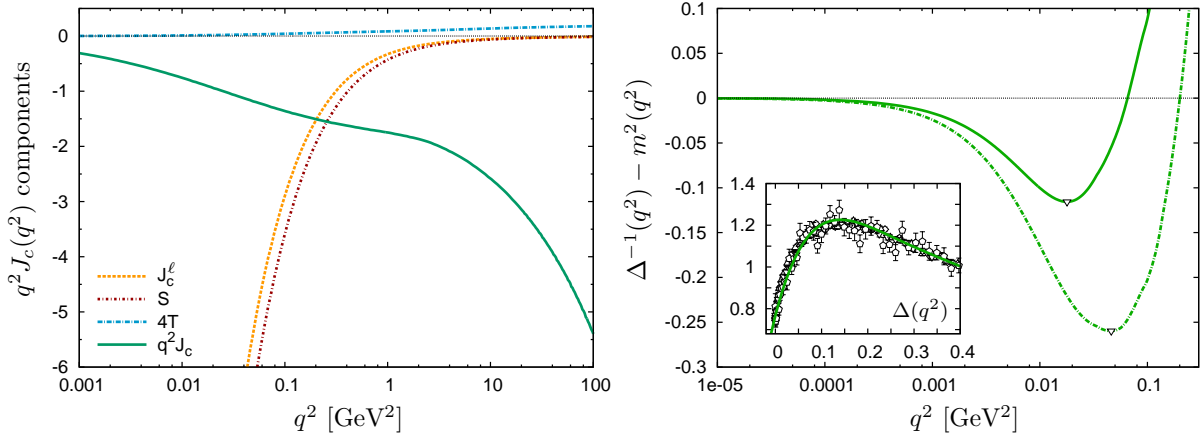}
\caption{\label{fig:q2Jc-lake-3d}(color online). The ghost kinetic term $q^2J_c(q^2)$ (left panel), and the propagator's full kinetic part $\Delta^{-1}(q^2)-m^2(q^2)$ (right panel), for the SU(2) three dimensional case. 
The two curves in the right panel corresponds to two different runnings for the dynamical mass.
Finally, the inset in the same panel shows the IR propagator fit together with the corresponding lattice data. }
\end{figure}

\begin{figure}[!t]
\includegraphics[scale=0.9]{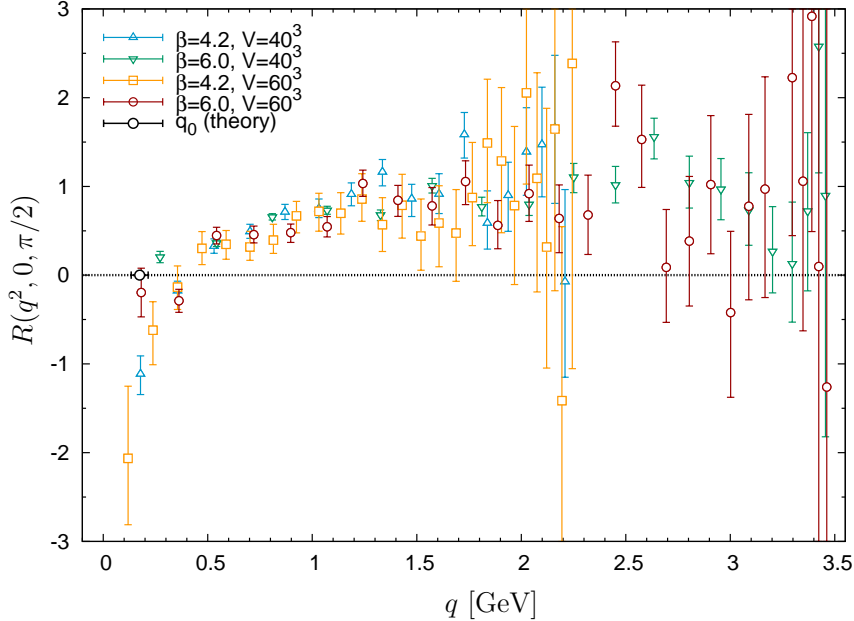}  
\caption{\label{fig:3gluon-3d}(color online). Prediction for the zero-crossing of the SU(2) form factor $R(q^2,0,\pi/2)$, measured on the lattice in three dimensions. The point shown is the average of the minima of the kinetic term for the two different runnings of the dynamical mass, with the error corresponding to the semidifference of these two points.}
\end{figure}

We next turn to the three dimensional case.  
In the  left panel of~\fig{fig:q2Jc-lake-3d} we plot the ghost kinetic term, 
showing the expected linear divergence; the latter translates into the well-known peak structure displayed 
in the IR by the gluon propagator (right panel inset)~\cite{Cucchieri:2010xr}. 
However, in order to repeat in $d=3$ the exercise of obtaining the 
kinetic term $q^2J(q^2)$ from~\1eq{ind-kinetic},
an additional assumption is needed. Specifically, 
 whereas the mass equation~\noeq{masseq} 
is valid also in $d=3$, the term~$\Y$ has not been computed in this case; as a consequence,  
no solutions for $m^2(q^2)$ are available. We will circumvent this difficulty by simply 
{\it assuming} that the three-dimensional gluon mass behaves in a way similar to that of $d=4$. 
Given the need for this additional assumption, it is natural to 
consider the corresponding results for $q^2J(q^2)$ as less definite than in the $d=4$ case.
In fact, in order to acquire a quantitative notion 
of how the running of the mass influences these results, 
we carry out the analysis twice, once for  
the four-dimensional $m^2(q^2)$ obtained for SU(2), and once for the corresponding mass in SU(3). 
The results are 
shown in the right panel of ~\fig{fig:q2Jc-lake-3d}.
Depending on the mass solution used, the position of the minimum 
of the kinetic term is located between $q_\s0\approx134$ (SU(2) mass) and $q_\s0\approx215$ MeV (SU(3) mass); 
this compares relatively well with lattice simulations of $R$~(\fig{fig:3gluon-3d}), which locate the zero-crossing at around $380$ MeV.

\section{\label{concl}Conclusions}

In this  work we have presented a  set of connections that link the IR
behavior  of  the  gluon  two-  and  three-point  sector  in  quenched
QCD. Specifically,  we have shown that  the fact that  the ghost field
remains  nonperturbatively massless,  as opposed  to the  gluon which
acquires  a  dynamically   generated  mass,  implies  unavoidably  the
existence of a negative IR divergence in the 
dimensionless co-factor  $J(q^2)$ of the  
kinetic part of the gluon
propagator (in $d=4$, $J\sim\ln q^2$, and in $d=3$, $J\sim1/q$).
This divergence, originating exclusively from the
one-loop  dressed diagrams  involving a  ghost loop,  does  not affect
the finiteness of the gluon two-point function,  since the  full kinetic  
term  is multiplied by a $q^2$.  
However, its presence manifests itself in at
least three  different ways: first, the dimensionful kinetic part $q^2 J(q^2)$
has a minimum, located at $q_{\s J}$; 
second, the full gluon  propagator $\Delta(q^2)$ displays a maximum, 
at a point denoted by  $q_{\s \Delta}$; 
third, a  (negative) divergence emerges in certain kinematic limits of  
the three-gluon  sector, where the standard lattice projector, $R(q^2)$, is proportional to $J(q^2)$;  
the point where $R(q^2)$ vanishes is denoted by $q_{\s 0}$. 

The PT-BFM formalism turns out to be particularly suited for  
verifying  the above picture quantitatively, mainly because 
it  allows for a  gauge-invariant  separation of  the ghost  and gluon
contributions to the gluon  propagator.
Consequently, one can identify
the divergent  ghost term  in a meaningful and unambiguous way. 
Specifically, the fact that the special ghost-gluon vertex $\widetilde{\Gamma}_\mu$ 
satisfies the QED-like WI of \1eq{WIGG}, furnishes 
a closed all-order expression for its longitudinal part, which is not possible to 
obtain for the corresponding vertex of the conventional ($R_{\xi}$) formulation.
In addition, when one combines the aforementioned feature of individual transversality
with the gluon mass equation and  
the available large-volume  lattice data, 
one is able to separate gauge-invariantly  the gluon- and ghost-loop contributions to the full kinetic term $q^2 J(q^2)$.

An additional interesting result in this context is the inequality between the special points  
$q_{\s J}$ and $q_{\s \Delta}$, namely $q_{\s J} < q_{\s \Delta}$, which constitutes a 
a clear and definite prediction of this particular approach based on the gluon mass generation.
Specifically, the aforementioned relation is a direct result of \1eq{dernonp}, and 
in particular of the monotonically decreasing nature of the gluon mass $m^2(q^2)$, as obtained from the 
corresponding dynamical equation. This relation, in conjunction with the  
indirect determination of $q^2 J_g(q^2)$
may provide valuable guidance in the 
effort to obtain the entire kinetic term of the gluon propagator from a 
complete treatment of the corresponding SDE.

\begin{figure}[!t]
\includegraphics[scale=.7]{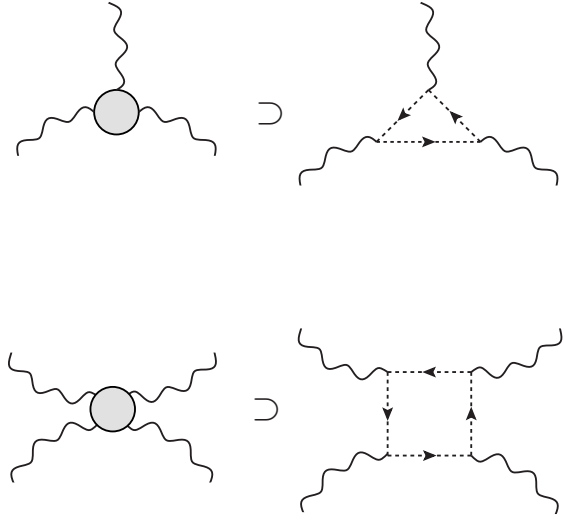}  
\caption{\label{3-and-4-gluon}The lowest order diagrams displaying a ghost loop in the case of the three- and four-gluon sector.}
\end{figure}

In the study of the $R$-projector, 
we have used as reference the results obtained for the vertex 
with three incoming background fields ($B^3$).
The main reason for this choice is the simplifications obtained 
due to the Abelian-like WI satisfied by the $B^3$ vertex. This property, in turn, 
eliminates all complications related to the ghost-gluon kernel, thus exposing the 
essence of the basic effect. For a very particular kinematic configuration
$R$ is expressed solely in terms of the kinetic term of the $B^2$ propagator,
a fact that imposes an exact coincidence between ${\widehat q}_{\s 0}$ and ${\widehat q}_{\s J}$.

Then, the corresponding results for the conventional vertex ($Q^3$) have been expressed 
as deviations from this prototypical case. Even though no exact results may be derived 
due to the ``contamination'' from the ghost-gluon vertex,   
the leading IR behavior can be accurately determined. On the other hand, the location of  ${q}_{\s 0}$ 
is not possible to pin down; however, it is reasonable to 
expect it to be relatively close to the 
corresponding point obtained for $B^3$.     
Actually, this value compares rather well with  the lattice data in $d=3$;
unfortunately, a similar comparison in 
$d=4$ is practically unattainable, since the corresponding 
lattice  simulations have not as yet firmly evidenced a sign change in $R$.

In this respect, it should be  noticed that, if taken at face value, our
results predict  that the lattice  volumes required in order to observe 
this  zero-crossing point are definitely large. 
This is a consequence of the fact that the
divergence in $J$  is only logarithmic in $d=4$,  something that pushes
the zero  crossing further into the  IR, when compared to  the $d=3$ case
(where the divergence in $J$  is linear).  Indeed, even if one assumes
a factor of 2 inaccuracy in the determination of $q_0$, one would still
need  $L\sim65$ (at  $\beta=2.5$ and  $a\approx21$ fm)  for  the SU(2)
gauge  group, and $L\sim30$  (at $\beta=5.7$  and $a\approx17$  fm) for
SU(3). It would seem, therefore, 
that SU(3) lattice  simulations would offer
better prospects in identifying the divergence in the $R$-projector.

Our  analysis suggests  that the sort of IR divergence 
considered here is likely to appear in other
Green's  functions  that contain  a  ghost  loop  at lowest  order  in
perturbation theory~(\fig{3-and-4-gluon}). 
In fact, the four-gluon vertex~\cite{Pascual:1980yu,Papavassiliou:1992ia,Ahmadiniaz:2013rla}, which
constitutes one  of the important missing ingredients  in the various
SDE studies, is a prime candidate for having such a divergence, due to
the (one-loop)  box-like  ghost diagram. On the  other hand,  Green's
function containing  (at least) one  external ${\bar c} c$  pair, (for
example, the ghost-gluon vertex) cannot have such a graph at one loop;
this type of graphs  appear at higher  orders, and the  additional loop integrations  
are expected to smoothen out the original divergence. It would be interesting to 
test the above conjectures by means of detailed calculations.

Since the origin of the effects described above  
is exclusively related to the presence of  
massless ghost  loops, the act of  ``unquenching'' not expected to modify 
our  results in  a  significant way.  Indeed,  large-volume  lattice
simulations~\cite{Ayala:2012pb}    and     the    corresponding    SDE
analysis~\cite{Aguilar:2012rz,Aguilar:2013hoa}  have  explicitly shown
that even  when dynamical quarks  are present  \n{i} the  ghost remains
massless,  and  \n{ii}  the  gluon acquires  dynamically  a  (heavier)
mass. As a result, the machinery developed  here is directly applicable  to the 
unquenched case, with the minimal modification
$\Delta^{-1}_\s{\mathrm{Q}}(q^2)-m^2_\s{\mathrm{Q}}(q^2)=q^2(J_c+J_g+J_q)$,
where   $\Delta^{-1}_\s{\mathrm{Q}}(q^2)$   and   $m^2_\s{\mathrm{Q}}$
represent,  respectively,  the   unquenched inverse propagator  and
dynamical mass, while $q^2J_q$ is the (IR finite) quark loop evaluated
in~\cite{Aguilar:2012rz,Aguilar:2013hoa}.

\acknowledgments 

The research of J.~P. is supported by the Spanish MEYC under 
grant FPA2011-23596. The work of  A.~C.~A  is supported by the 
National Council for Scientific and Technological Development - CNPq
under the grant 306537/2012-5 and project 473260/2012-3,
and by S\~ao Paulo Research Foundation - FAPESP through the project 2012/15643-1. 

\appendix

\section{\label{appA}The vertex $B^3$ and its $R$-projector}

The $B^3$ vertex  
exposes the basic divergent features of the $R$-projector, without the additional complications of the conventional vertex.
In this Appendix we present some of the relevant technical points in this context.

At tree-level, the $B^3$ and the conventional ($Q^3$) vertices coincide, 
\be
 \widehat\Gamma_{\alpha\mu\nu}^{(0)}(q,r,p) = \Gamma_{\alpha\mu\nu}^{(0)}(q,r,p) =
(q-r)_{\nu}g_{\alpha\mu} + (r-p)_{\alpha}g_{\mu\nu} + (p-q)_{\mu} g_{\alpha\nu},
\ee
where all momenta are entering. 
Beyond tree-level the two vertices differ, and are related 
by a complicated all-order BQI; in addition, both vertices are completely Bose-symmetric.

More important in the present context is the fact that 
that $\widehat{\Gamma}$ satisfies Abelian WIs; one has
\be
q^\alpha\widehat\Gamma_{\alpha\mu\nu}(q,r,p) = p^2 {\widehat J} (p^2)P_{\mu\nu}(p)-r^2 {\widehat J}(r^2)P_{\mu\nu}(r),
\label{BBB-WI}
\ee
with 
\be
J(q^2)=F^2(q^2){\widehat J}(q^2);  
\label{JhatJ}
\ee
cyclic permutations of indices and momenta generate the remaining WIs.

These identities are to be contrasted with the STIs satisfied by the conventional vertex 
\be 
q^\alpha\Gamma_{\alpha\mu\nu}(q,r,p)=F(q^2)\left[p^2 J(p^2)P_\nu^\alpha(p)H_{\alpha\mu}(p,q,r)-
r^2 J(r^2)P_\mu^\alpha(r) H_{\alpha\nu}(r,q,p)\right],
\label{QQQ-STI}
\ee
(and cyclic permutations), which explicitly involve the ghost-kernel~$H$.

The complete closed form of  $\widehat{\Gamma}$ is not known; its longitudinal part, however, 
may be reconstructed by `solving' the identities~\noeq{BBB-WI}~\cite{Ball:1980ax}.
Specifically, one begins by separating the vertex into the 
``longitudinal'' and the (totally) ``transverse'' parts,   
\be
\widehat\Gamma_{\alpha\mu\nu}(q,r,p)= 
\widehat\Gamma^{\ell}_{\alpha\mu\nu}(q,r,p) + \widehat\Gamma^{t}_{\alpha\mu\nu}(q,r,p), 
\label{decomp}
\ee
where the component $\widehat\Gamma^{\ell}$ satisfies the 
WIs of \1eq{BBB-WI} (and its permutations), whereas  
$q^{\alpha}\widehat\Gamma^{t}_{\alpha\mu\nu} = 
r^{\mu}\widehat\Gamma^{t}_{\alpha\mu\nu}=
p^{\nu} \widehat\Gamma^{t}_{\alpha\mu\nu}= 0$. 

The longitudinal part is then decomposed into 10 form factors, $\widehat{X}_i$, according to
\be
\widehat\Gamma^{\ell}_{\alpha\mu\nu}(q,r,p)=\sum_{i=1}^{10}\widehat{X}_i(q,r,p)\ell^i_{\alpha\mu\nu},
\label{tenlon}
\ee
with the explicit form of the tensors $\ell^i$ given by~\cite{Binger:2006sj}
\begin{align}
\ell^1_{\alpha\mu\nu} &=  (q-r)_{\nu} g_{\alpha\mu};& 
\ell^2_{\alpha\mu\nu} &=  - p_{\nu} g_{\alpha\mu};&
\ell^3_{\alpha\mu\nu} &=  (q-r)_{\nu}[q_{\mu} r_{\alpha} -  (q\spr r) g_{\alpha\mu}],&
\label{Ls}
\end{align}
with $\ell^{i+3}_{\alpha\mu\nu}$ given by cyclic permutations of momenta and indices and $\ell^{10}_{\alpha\mu\nu} = q_{\nu}r_{\alpha}p_{\mu} + q_{\mu}r_{\nu}p_{\alpha}$.

The WIs of \noeq{BBB-WI}  give rise to an algebraic system 
for the $\widehat{X}_i$, whose solution reads~\cite{Binger:2006sj}
\begin{align}
\widehat{X}_1 &= \frac{1}{2}[\widehat{J}(q^2) + \widehat{J}(r^2)];&\widehat{X}_2 &= \frac{1}{2}[\widehat{J}(q^2) - \widehat{J}(r^2)];&
\widehat{X}_3 &= \frac{\widehat{J}(q^2) - \widehat{J}(r^2)}{q^2 - r^2},
\label{factorsBBB}
\end{align}
with $\widehat{X}_{i+3}$ obtained from $\widehat{X}_{i}$ as before, and with $\widehat{X}_{10}=0$.

We thus see that the longitudinal form factors constituting  $\widehat\Gamma^\ell$ involve {\it only} the quantity $\widehat{J}$; instead,  as seen in~\2eqs{X7general}{X9general}, the corresponding expressions for the form factors of the conventional vertex,  contain, in addition, the ghost dressing function $F$ and the various form-factors comprising the gluon-ghost kernel $H$.

Finally, the (undetermined) transverse part of the vertex is described by 4 remaining form factors $\widehat{Y}_i$, 
\be
\widehat\Gamma^{t}_{\alpha\mu\nu}(q,r,p) = \sum_{i=1}^{4}\widehat{Y}_i(q,r,p)t^i_{\alpha\mu\nu},
\label{vtr}
\ee
with the completely transverse tensors $t^i$ given by
\be
t^1_{\alpha\mu\nu} =
[(q\spr r) g_{\alpha\mu} - q_{\mu}  r_{\alpha}]
[(r\spr p) q_{\nu} - (q\spr p) r_{\nu}],
\ee
$t^2_{\alpha\mu\nu}$ and $t^3_{\alpha\mu\nu}$ obtained from this expression by  cyclic permutations, and, finally,
\begin{align}
t^4_{\alpha\mu\nu}&=g_{\mu\nu}[ (p\spr q)r_\alpha-(r\spr q)p_\alpha ]+g_{\alpha\mu}[(r\spr p)q_\nu-(q\spr p)r_\nu ] +g_{\alpha\nu}[(r\spr q)p_\mu -(r\spr p)q_\mu]\nonumber \\
&+p_\alpha q_\mu r_\nu-r_\alpha p_\mu q_\nu.
\label{Ts}
\end{align}

Using these decompositions, it is straightforward to evaluate  the $R$-projector 
defined in~\1eq{Rdefinition}. In particular, one obtains (in $d$ dimensions)\footnote{Notice that the following expressions 
are general, and apply also to the conventional vertex case, 
with the obvious replacements $\widehat{X}_i\to X_i$, where the $X_i$ are now determined by the STIs~\noeq{QQQ-STI}.}~\cite{Binosi:2013rba}
\begin{align}
{\cal N}^{\ell}(q,r,p) &= 4\frac{r^2 p^2 - (r\spr p)^2}{q^2 r^2 p^2} \left\{ [(d-1)q^2r^2 - (q\spr p)(p\spr r)]\widehat{A}_1 + [(d-1)r^2p^2 - (p\spr q)(q\spr r)]\widehat{A}_2 \right.\nonumber \\
&+ \left.[(d-1)q^2p^2 - (q\spr r)(r\spr p)]\widehat{A}_3 + [(q\spr r)(r\spr p)(p\spr q) - q^2r^2p^2]\widehat{A}_4 \right\},\nonumber \\
{\cal N}^{t}(q,r,p) &= 2[r^2 p^2 - (r\spr p)^2] \left\{ [(d-1)(q\spr r)-p^2]\widehat{Y}_1 + [(d-1)(r\spr p)-q^2]\widehat{Y}_2\right.\nonumber \\
& +\left.[(d-1)(q\spr p)-r^2]\widehat{Y}_3 + 3(d-2)\widehat{Y}_4 \right\},
\end{align}
and
\begin{align}
{\cal D}(q,r,p)& = 4\frac{r^2 p^2 - (r\spr p)^2}{q^2 r^2 p^2} [(d-1)(q^2 r^2 + q^2 p^2 + r^2 p^2) + (r\spr p)^2 - r^2 p^2],
\end{align}
where we have defined ${\cal N}={\cal N}^\ell+{\cal N}^t$, while the combinations $\widehat{A}_i$ are given by
\begin{align}
\widehat{A}_1 &= \widehat{X}_1 - (q\spr r)\widehat{X}_3; & 
\widehat{A}_2 &= \widehat{X}_4 - (r\spr p)\widehat{X}_6; \nonumber \\
\widehat{A}_3 &= \widehat{X}_7 - (p\spr q)\widehat{X}_9; & 
\widehat{A}_4 &= -\widehat{X}_3 - \widehat{X}_6 - \widehat{X}_9.
\label{Afactors}
\end{align}

We next consider three particular kinematic configurations of the $\widehat R$-projector, 
which are typically simulated on the lattice~\cite{Cucchieri:2006tf,Cucchieri:2008qm}. 
Interestingly enough, as we will see, they all display the same exact divergent behavior in the IR.

\subsubsection{Orthogonal configuration with one momentum vanishing}

In this case we take $\varphi=\pi/2$ and $r\to 0$; as in the latter limit ${\cal N}^t$ vanishes, we obtain the simple result~\cite{Binosi:2013rba}
\be
\widehat R(q^2,0,\pi/2) =[q^2\widehat{J}(q^2)]'.
\label{hatRsim}
\ee
The above result is exact, and valid for any $q^2$; 
in particular, using \2eqs{JhatJ}{Jcdiv}, we find the leading IR behavior to be
\be
\widehat R(q^2,0,\pi/2) \underset{q^2\to0}{=}  C_d F^{-1}(0) \int_k\frac{F(k)}{k^2(k+q)^2}.
\label{Rfullapp}
\ee
If at this point we set $F=1$, and carry out the resulting (effectively one-loop) integral, we recover \1eq{IR-be}. 

It is interesting to observe that in this case, since~\1eq{hatRsim} is exact, $\widehat R^{\mathrm{s}\ell}$ vanishes identically, 
so that the condition~\1eq{Rqstar} simplifies to  
\be
\widehat R({\widehat q}_\s{\Delta}^2) = - [\widehat m^2({\widehat q}_\s{\Delta}^2)]^{\prime}.
\label{Rqstar-simp}
\ee
This means that for the equality $\widehat q_\s{\Delta}=\widehat q_\s0$ to hold, the mass $\widehat m^2(q^2) = F^{-2}(q^2)  m^2(q^2)$ should not be monotonic; while a preliminary study shows that this is indeed what happens, this issue needs to be thoroughly investigated. 

Note, finally, that the ratio between \1eq{Rdivfinal} and \1eq{Rfullapp} is finite, and given by   
\be
\frac{R (0)}{\widehat R (0)} = F^{3}(0),
\label{Rratios}
\ee
as was first derived in~\cite{Binosi:2013rba}, following different considerations.

\subsubsection{Orthogonal configuration with equal momenta}

In this case one has
\begin{align}
q^2=&r^2;&
q\spr r&=0;&
p^2&=2q^2;&
q\spr p&=r\spr p=-q^2;&
\varphi=\pi/2.
\end{align}
Clearly, in this configuration, the transverse part of the vertex survives,  
\be
\frac{{\cal N}^{t}(q^2,q^2,\pi/2)}{{\cal D}(q^2,q^2,\pi/2)}=
\frac{q^2}{5d-6}\left[2q^2\widehat{Y}_1+dq^2(\widehat{Y}_2+\widehat{Y}_3)-3(d-2)\widehat{Y}_4\right],
\ee
where $\widehat{Y}_i=\widehat{Y}_i(q^2,q^2,\pi/2)$. However, if we assume that the form-factors $\widehat{Y}_i$ do not 
contain poles in $q^2$, it is clear that this term vanishes in the IR. 
Then, from ${\cal N}^{\ell}$ we obtain
\be
\widehat{R}(q^2,q^2,\pi/2)=\frac1{5d-6}\left[2q^2\widehat{J}\,'(q^2)-(d+4)\widehat{J}(q^2)+2(3d-1)\widehat{J}(2q^2)\right] +\cdots,
\ee
where the omitted terms are subleading in the IR. This result may be easily rearranged to read 
\be
\widehat{R}(q^2,q^2,\pi/2)= \widehat{J}(q^2) + 
\frac1{5d-6} \left\{2q^2\widehat{J}\,'(q^2) + 2(3d-1) \left[\widehat{J}(2q^2) - \widehat{J}(q^2) \right]\right\},
\ee
where the first term is leading, and coincides with that of \1eq{Rfullapp}, while the second is subleading.

\subsubsection{All momenta equal}

Setting
\begin{align}
q^2&=r^2=p^2;& q\spr r&=q\spr p=p\spr r=-\frac{q^2}2;&
\varphi=2\pi/3,
\end{align}
we find for the transverse part 
\be
\frac{{\cal N}^{t}(q^2,q^2,2\pi/3)}{{\cal D}(q^2,q^2,2\pi/3)}=
\frac{q^2}{12d-15}\left[(d+1)q^2(\widehat{Y}_1+\widehat{Y}_2+\widehat{Y}_3)-6(d-2)\widehat{Y}_4\right],
\ee
where $\widehat{Y}_i=\widehat{Y}_i(q^2,q^2,2\pi/3)$. Again, this term may be neglected under the same assumptions stated above.
Then, one finds
\be
\widehat{R}(q^2,q^2,2\pi/3)= \widehat{J}(q^2) + \frac{2(d-1)}{4d-5} q^2\widehat{J}\,'(q^2),
\ee
thus obtaining exactly the same leading IR behavior as in the previous two cases.

\section{\label{appB}The transverse part of the vertex $\widetilde{\Gamma}_{\mu}$}

The contribution of the transverse part of $\widetilde{\Gamma}_{\mu}$ to $J_c(q^2)$, to be denoted by  $J_{c}^{t}(q^2)$,
is given, up to irrelevant constants and the finite ghost dressing function $F(q)$, by 
\bea
q^2 J_{c}^{t}(q^2) &\sim & \int_k k^{\mu} D(k) D(k+q) \left\{(k\cd q)(k+q)_\mu- \left[(k+q)\cd q)\right]k_\mu\right\} {\cal A}(k,k+q)
\nonumber\\
&\sim & \int_k D(k) D(k+q){\cal A}(k,k+q) \left[(k\cd q)^2 - q^2 k^2\right].
\eea
So, after passing to spherical coordinates, and using that  $(k\cd q)^2 = q^2 k^2 \cos^2\theta$, we obtain 
\be
J_{c}^{t}(q^2)  \sim \int_k D(k) D(k+q){\cal A}(k,k+q) k^2 \sin^2\theta.
\ee
Now, at $q=0$, we have (setting $k^2 =y$)
\be
J_{c}^{t}(0)  \sim \int_k   D^2(y) {\cal A}(y) y\, \sin^2\theta,
\ee
with 
\be
\int_k\ =\ \frac1{(2\pi)^d}\frac{\pi^{\frac{d-1}2}}{\Gamma\left
(\frac{d-1}2\right)}\int_0^\pi\! \mathrm{d}\theta\sin^{d-2}\theta\int_0^\infty\! \mathrm{d}y\,y^{\frac d2-1}.
\label{d-measure}
\ee
The integration over the angle $\theta$ furnishes an additional irrelevant constant, 
and so, after using \1eq{ghostdress}, 
\be
J_{c}^{t}(0) \sim \int_0^\infty\! \mathrm{d}y\, y^{\frac d2-2} F^2(y){\cal A}(y).
\ee
Thus, if we assume that, in the deep IR, ${\cal A}(y) \sim y^{a}$, then 
the lower limit of this integral is finite provided that $a> 1- d/2$ 
(remember that $F(y)$ saturates to a constant in the IR).
Thus, for $d=4$, one gets a finite (and, therefore, subleading) contribution to $J_c(q^2)$, provided 
that ${\cal A}(y)$ diverges weaker than a simple pole.
In $d=3$, the corresponding limiting case is a square root of a pole.
 
The above  
conditions appear relatively easy to satisfy, especially since the diagrammatic 
representation of $\widetilde{\Gamma}_{\mu}$ does not include a ghost-loop at lowest order,
according to the discussion presented in Section~\ref{concl}.


\end{document}